\documentclass[onecolumn]{aastex631}

\usepackage{amsmath}
\DeclareUnicodeCharacter{2212}{-}
\graphicspath{{./}{figures/}}
 
\shorttitle{Pulsating WC-late type star}
\shortauthors{Kar et al.}

\begin{document}

\title{Detection of high-frequency pulsation in WR\,135: investigation of stellar wind dynamics}

\author[0000-0001-7874-0218]{Subhajit Kar}
\affil{S.N. Bose National Centre for Basic Sciences,
Kolkata - 700106, India}
\correspondingauthor{Subhajit Kar}
\email{subhajit0596@gmail.com, subhajitksnbac@bose.res.in}

\author[0000-0002-5440-7186]{Ramkrishna Das}
\affiliation{S.N. Bose National Centre for Basic Sciences,
Kolkata - 700106, India}

\author[0000-0002-7254-191X]{Blesson Mathew}
\affiliation{Department of Physics and Electronics, Christ University, Bangalore - 560029, India.}

\author[0000-0003-0295-6586]{Tapas Baug}
\affiliation{S.N. Bose National Centre for Basic Sciences,
Kolkata - 700106, India}

\author[0009-0004-0275-5201]{Avijit Mandal}
\affiliation{S.N. Bose National Centre for Basic Sciences,
Kolkata - 700106, India}

\begin{abstract}

We report the detection of high-frequency pulsations in WR\,135 from short cadence (10\,minutes) optical photometric and spectroscopic time series surveys. The harmonics up to $6^{th}$ order are detected from the integrated photometric flux variations while the comparatively weaker $8^{th}$ harmonic is detected from the strengths of the emission lines. We investigate the driving source of the stratified winds of WR\,135 using the radiative transfer modeling code, CMFGEN, and find the physical conditions that can explain the propagation of such pulsations. From our study, we find that the optically thick sub-sonic layers of the atmosphere are close to the Eddington limit and are launched by the Fe-opacity. The outer optically thin super-sonic winds ($\tau_{ross}=0.1-0.01$) are launched by the He\,$\textsc{ii}$ and C\,$\textsc{iv}$ opacities. The stratified winds above the sonic point undergo velocity perturbation that can lead to clumps. In the optically thin supersonic winds, dense clumps of smaller size ($f_{VFF}=0.27-0.3$, where $f_{VFF}$ is the volume filling factor) pulsate with higher-order harmonics. The larger clumps ($f_{VFF}=0.2$) oscillate with lower-order harmonics of the pulsation and affect the overall wind variability.

\end{abstract}


\section{Introduction}\label{sect:intro}
Classical Wolf-Rayet (cWR) stars are the Hydrogen-depleted class of WR stars currently undergoing core He-burning. Based upon the stage of their evolution, the cWR stars can be classified as WN (quantified by helium and nitrogen lines in the spectrum), WC (helium and carbon lines), or WO (carbon and oxygen lines) types, that are further sub-categorized as WR-Early and WR-Late types based on their effective surface temperatures. In general, WR stars are identified from their strong and broad emission line profiles reflecting supersonic ($\geq$1000\,km\,$\mathrm{s^{-1}}$) stellar outflows or winds. Driven by the intense UV radiation, stellar winds profoundly influence stellar evolution by carrying away large amounts of stellar material ($\dot{M}\sim10^{-4}-10^{-5}\,M_{\odot}yr^{-1}$) that simultaneously enrich the surrounding Interstellar medium (ISM). cWR stars are found to be spectroscopically and polarimetrically variable due to the inhomogeneous and asymmetric nature of the winds. Most of the cWR stars show intrinsic variability either due to stellar rotations, photospheric pulsations, or wind inhomogeneities. Rotation leads to the motion of the corotating interaction regions (CIRs) \citep{2009ApJ...698.1951S, 2011ApJ...736..140C} while the stellar pulsations and wind inhomogeneities are caused by perturbations generated at the base of the stellar winds due to instabilities \citep{2009A&A...499..279C}. Wind inhomogeneities being stochastic in nature are detected from the non-periodic radially outward displacement of the line subpeaks from the line center to the scattering wings. This occurs due to the motion of discrete wind emission elements \citep{1999ApJ...514..909L, 2000AJ....120.3201L} or small, inhomogeneous density structures in stellar winds \citep{1988ApJ...334.1038M}, which lead to small-scale variability \citep{2020ApJ...903..113C} in the spectral lines ($\sigma<$ 5\,\%). The CIRs identified from the periodic epoch-wise (few days) displacements of subpeaks across the primary broad line profile show large-scale variability ($\sigma>$ 5\,\%). Studying these two types of wind structures can constrain the mass loss rates that affect the later stages of their evolution.

A long-standing problem in WR stars is to understand the driving source behind the mass outflow in cWR stars. 
It has been found that WR stars can undergo envelope inflation \citep{2012A&A...538A..40G} due to strong opacity from the Fe-M shell (i.e., Fe\,$\textsc{ix}$-$\textsc{xvi}$) transitions \citep{2020MNRAS.491.4406S} which not only drive the inner winds but also sometimes generate strange mode instabilities (SMIs, \citet{1994MNRAS.271...66G, 2008ASPC..391..307G}) that are responsible for the inhomogeneities in the stellar winds as well as the excitement of strange-mode pulsations (SMPs) in some cases. In the radiation-dominated atmospheres of WR stars, energy transport mainly occurs due to diffusion. This gives rise to phase shifts between the pressure and density perturbations that are responsible for the existence of damped and excited modes occurring in complex conjugate pairs, referred to as strange modes. The position of the opacity ($\kappa$) maxima determines the frequency of the SMPs  \citep{2009CoAst.158..245S}, while the SMIs are independent of $\kappa$-mechanism. From a non-linear perspective, \citet{2008ASPC..391..307G} showed that these pulsations can achieve longer periods (P$\sim$10\,hours) due to shock-induced inflations from the SMIs. In contrast, \citet{1988ApJ...335..914O} derived an alternative solution to explain pulsations and stochastic variability in line-driven stellar winds. The line-deshadowing instabilities (hereafter, LDIs) generated at the base of the winds \citep{2018A&A...611A..17S} due to self-excited radiative acoustic waves \citep{1980ApJ...242.1183A} lead to inhomogeneities in the stellar winds. Such instabilities can also trigger long-period non-radial pulsations in the sub-sonic winds that are amplified in the supersonic winds by the LDIs present in those winds.  

\citet{2022ApJ...925...79L} found that most of the cWR stars showed intrinsic variability due to the presence of stochastic inhomogeneities in the stellar winds. Moreover, \citet{2020ApJ...903..113C} noticed that the cooler cWR stars with slower wind velocity ($<2000$ $\mathrm{kms^{-1}}$) are highly variable than the hotter stars with faster wind velocities. Among WC-type stars, the WCL-type stars are the cooler subtypes known for higher mass-loss rates than the WCE-type stars. WCL-type (WC7-9) stars exhibit line profile variations (extrinsic variability) often because of a binary companion but across a much longer timescale (in years, \citet{2020MNRAS.497.4448S}) than the stochastic variability (in hours, \citet{2022ApJ...925...79L}).  

Till now, a few WCL-type stars have simultaneously shown both stochastic as well as pulsational variability. Intensive monitoring of C\,$\textsc{iii}$ $\lambda$5696 and C\,$\textsc{iv}$ $\lambda\lambda$5802-12 emission lines of WR\,135 (a single WC8-type star) by \citet{2000AJ....120.3201L} revealed that this object has small-scale structures in its stellar winds. Later studies \citep{2020ApJ...903..113C, 2022ApJ...925...79L} found that WR\,135 exhibits characteristic small-scale stochastic phenomena formed as a consequence of the clumpy winds \citep{1988ApJ...335..914O}. Photometric survey \citep{2021MNRAS.502.5038N} of WR\,135 classified it as a pulsating WR star.
Therefore, WR\,135 shows both stochastic and pulsational variability likely due to SMI-induced wind clumping and SMPs, respectively. In this study, we derive the driving mechanism of the stellar winds of WR\,135 that can support both these phenomena simultaneously. We also search for pulsational variability from the varying strengths of the unblended emission lines present in a short cadence spectroscopic time series dataset (Sec. \ref{sect:analysis}). We further analyze the photometric time series data to identify the higher-order harmonics.  In Sec. \ref{subsec:model}, we look at the impact of velocity gradient and the clumping factor on the line formation in the optically thin regions of the supersonic winds. In Sec. \ref{subsec:driving}, we determine the source of the wind driving and discuss the physical conditions that can simultaneously support different harmonics of the pulsation in Sec. \ref{sect:discussion}. The work is concluded in Sec. \ref{sect:conclusion}.

\section{Data and Analysis} \label{sect:Obs}

\subsection{Photometric data}\label{subsec:photometric_data}
\begin{figure*}
   \centering
   \includegraphics[width=0.8\textwidth]{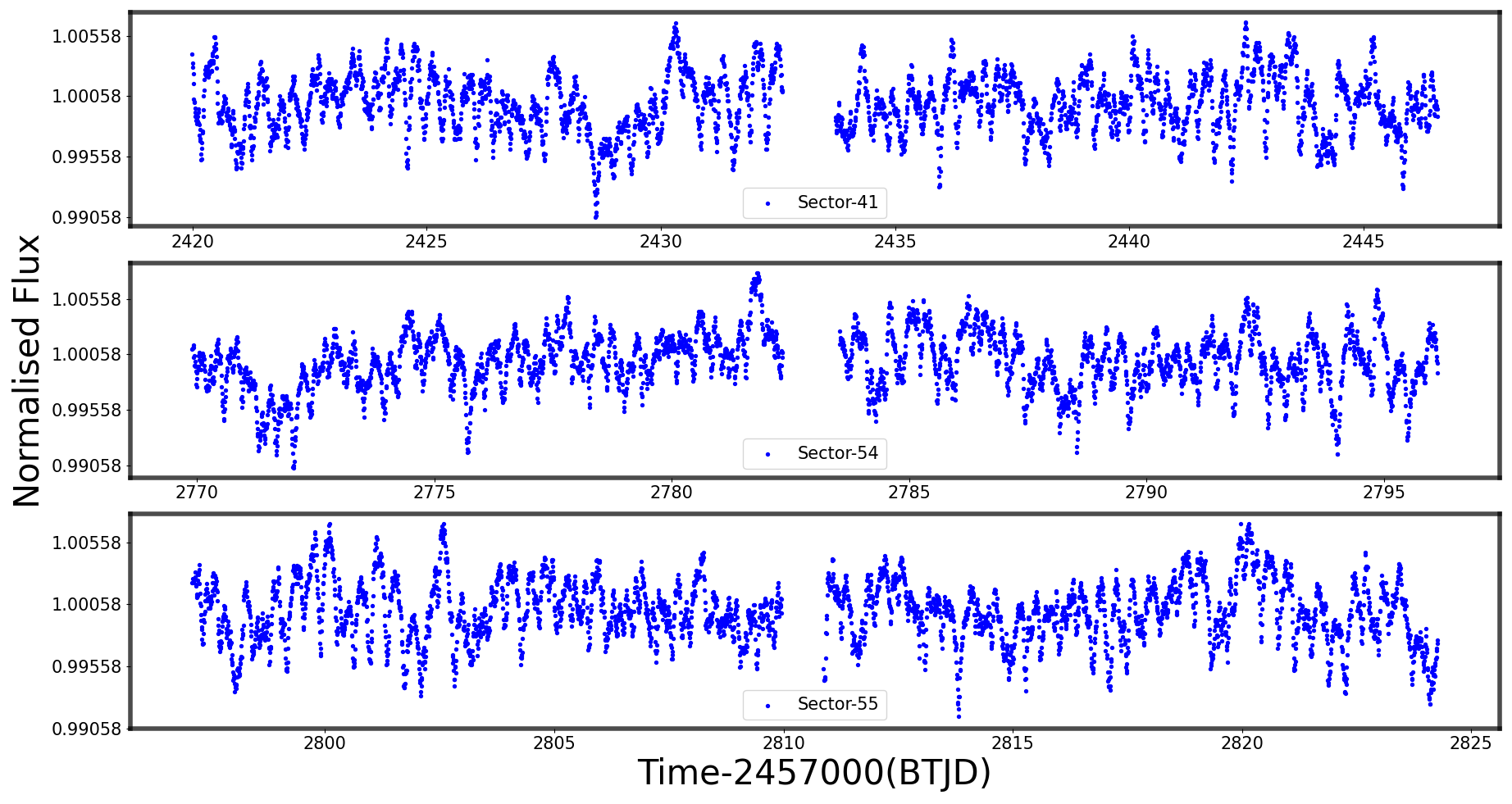}
   \caption{Normalised lightcurves of WR\,135 with 10\,minutes cadence derived from FFIs observed in Sector-41, 54 \& 55.}
   \label{fig:TESS_lightcurve}
\end{figure*}

We used optical (600$-$1000\,nm) photometric data observed by the Transiting Exoplanet Survey Satellite (TESS), which is a state-of-the-art space telescope built to detect exoplanets and asteroseismic phenomena \citep{10.1117/1.JATIS.1.1.014003} with a Field-Of-View (FOV) of $\mathrm{96^{\circ}\times24^{\circ}}$. It captures focussed data of pre-selected objects at 2\,minutes cadence while the Full Frame Images (FFIs) covering the total FOV are captured at longer cadences (30\,minutes during the TESS Prime mission and 10\,minutes during the first Extended Mission). During each orbit, the TESS scans a particular patch (aka Sector) of the sky for $\sim$27 consecutive days. The precise detection capability of TESS makes it an ideal telescope to monitor the intrinsic variability of massive stars (e.g. WR stars) across different timescales (hours to days). The TESS photometric time series data (with 10\,minutes cadence) of WR\,135 were observed in Sectors 41, 54, and 55. Utilizing the MIT Quick Look Pipeline \citep{2020RNAAS...4..204H, Kunimoto_2021}, \citet{2022RNAAS...6..236K} extracted the lightcurves from the 10\,minute FFIs using an optimal aperture. Following the Presearch Data Conditioning algorithm \citep{Smith_2012, Stumpe_2012}, the lightcurves were post-processed to remove outliers and trends due to instrumental noise. Finally, the lightcurves were normalized with the mean flux. The flux-normalized lightcurves are publicly available as High-Level Science Products (HLSP) in the Mikulski Archive for Space Telescopes (MAST). We retrieved and analyzed the data using the $\textit{Lightkurve}$ package \citep{2018ascl.soft12013L}. The corresponding light curves are shown in Fig. \ref{fig:TESS_lightcurve}.

We also used the optical and Infrared (IR) photometric data (Integrated Flux density) from the online accessible Vizier data catalogue\footnote{\url{https://vizier.cds.unistra.fr/}}.  Table \ref{tab:phot} shows the photometric data from optical to mid-IR bands along with the respective catalogs. Optical B and V-band flux data were obtained from the data releases by \citet{1997yCat.1239....0E} and \citet{2008AJ....136..735L} respectively. Optical G-band data were acquired from GAIA DR2 data release \citep{2018A&A...616A...1G}. Near-infrared (NIR) data from the 2MASS data release \citep{2006AJ....131.1163S} were acquired. Additionally, mid-infrared (mid-IR) data were sourced from WISE (Wide-field Infrared Survey Explorer) data release \citep{2012wise.rept....1C}, and SASS (Spitzer Atlas of Stellar Spectra) data release \citep{2010ApJS..191..301A}.

\begin{deluxetable*}{cccc}
\tablecaption{\label{tab:phot}Integrated flux data of WR\,135 across optical-IR bands.}
\tablewidth{0pt}
\tablehead{
\colhead{Band} & \colhead{$\lambda_{c}$} & \colhead{Flux density} & \colhead{Data Catalogue} \\
\colhead{} & \colhead{(\AA)} & \colhead{(erg\,s$^{-1}$\,cm$^{-2}$\,\AA$^{-1}$)} & \colhead{}
}
\startdata
B & 4440 & 3.01$e^{-12}$ & \citet{2007yCat.1305....0L} \\
V & 5540 & 2.03$e^{-12}$ & \citet{1997yCat.1239....0E} \\ 
G  & 6230 & 1.79$e^{-12}$ & \citet{2018yCat.1345....0G} \\
J & 12400 & 3.94$e^{-13}$ & \citet{vizierII246} \\
H & 16500 & 1.65$e^{-13}$ & " \\
$K_{s}$ & 21600 & 9.35$e^{-14}$ & " \\
W1 & 33500 & 2.26$e^{-14}$ & \citet{2012yCat.2311....0C} \\
W2 & 46000 & 9.45$e^{-15}$ & " \\
IRAC-8.0 & 78700 & 1.63$e^{-15}$ & \citet{2010yCat..21910301A} \\
W3 & 116000 & 4.8$e^{-16}$ & \citet{2012yCat.2311....0C} \\
W4 & 221000 & 7.74$e^{-17}$ & " \\
\enddata
\end{deluxetable*}

\subsection{Spectroscopic data}\label{subsec:spectroscopic_data}
\subsubsection{3.6m-Canadian French Hawaii Telescope}\label{subsubsec:cfht}
High-resolution (R$\sim$68000) optical spectroscopic time-series data (with SNR$\sim300-400\,pixel^{-1}$) of WR\,135 were retrieved from the Canadian Astronomy Data Centre\footnote{\url{https://www.cadc-ccda.hia-iha.nrc-cnrc.gc.ca/en/}}(CADC). The spectropolarimetric data were observed by \citet{2014ApJ...781...73D} using the fibre-fed ESPaDOnS\footnote{\url{http://www.cfht.hawaii.edu/Instruments/Spectroscopy/Espadons/}} mounted at the 3.6m-Canadian French Hawaii Telescope (CFHT). The spectrograph covers the entire optical band (3500$-$10000\,\AA) across 40 spectral orders in a single exposure making it suitable for simultaneous monitoring of different spectroscopic features over the broad wavelength region. The Echelle spectroscopic data of WR\,135 were reduced and wavelength-calibrated using the Libre-ESpRIT \citep{1997MNRAS.291..658D} based Upena software as discussed in \citet{2014ApJ...781...73D}. A polynomial was fitted to the spectral continuum chosen from three regions with less or no prominent spectral features (i.e. pseudo-continuum). Finally, the polynomial-fitted continuum was used to normalize the spectra used in this study. The log of the observations is shown in Table \ref{tab:observation}.

For our analysis, we extract the spectral region with emission lines using the $\textit{astropy}$\footnote{\url{https://www.astropy.org/}} \citep{2013A&A...558A..33A, 2018AJ....156..123A, 2022ApJ...935..167A}. We utilize the $\textit{extract\_region}$ task and fit individual line profiles with a 1-D Gaussian function ($\textit{fit\_lines}$ task in the $\textit{fitting}$ module). The best fitting profile is determined from the Levenberg Marquardt chi$-$square minimization \citep{Levenberg1944AMF, doi:10.1137/0111030}. The equivalent widths (EWs) of the spectral lines are then derived from the fitted model. The errors in the EWs of the emission lines are estimated using the method discussed in \citet{2006AN....327..862V}. The error depends on the ratio of flux associated with the emission line and adjacent continuum and is inversely proportional to the SNR of the same continuum region. Therefore, we compute the normalized mean-flux for the line ($\overline{F_{line}}$) and continuum ($\overline{F_{cont}}$) from a featureless spectral region adjacent to the emission line. The same pseudo-continuum region is used to estimate the signal-to-noise ratio (SNR) of the spectra.

\begin{deluxetable*}{ccccccc}
\tablecaption{Log of spectroscopic observations for WR\,135 \label{tab:observation}}
\tablewidth{0pt}
\tablehead{
\colhead{Date} & \colhead{Epoch} & \colhead{Telescope} & \colhead{Instrument} & \colhead{Resolution} & \colhead{$\lambda$} & \colhead{Exposure} \\
\colhead{} & \colhead{(JD 240000+)} & \colhead{} & \colhead{} & \colhead{} & \colhead{(\AA)} & \colhead{}
}
\startdata
2008 June 28  & 54645.96573 $-$ & CFHT (archival) & ESPaDOnS & 68000 & 3000-10000 & 12\,frames * 600\,s\\ 
 & 54646.05832 & & & & & \\ \hline
2023 September 19  & 60207.29852 & HCT & HFOSC/Gr7 & 1400 & 3800$-$7000 & 1\,frame * 15\,s\\
"  & 60207.30182 & " & HFOSC/Gr8 & 2200 & 5000$-$9000 & 1\,frame * 60\,s\\
\enddata
\end{deluxetable*}

\begin{figure}
    \centering
    \includegraphics[width=0.8\textwidth]{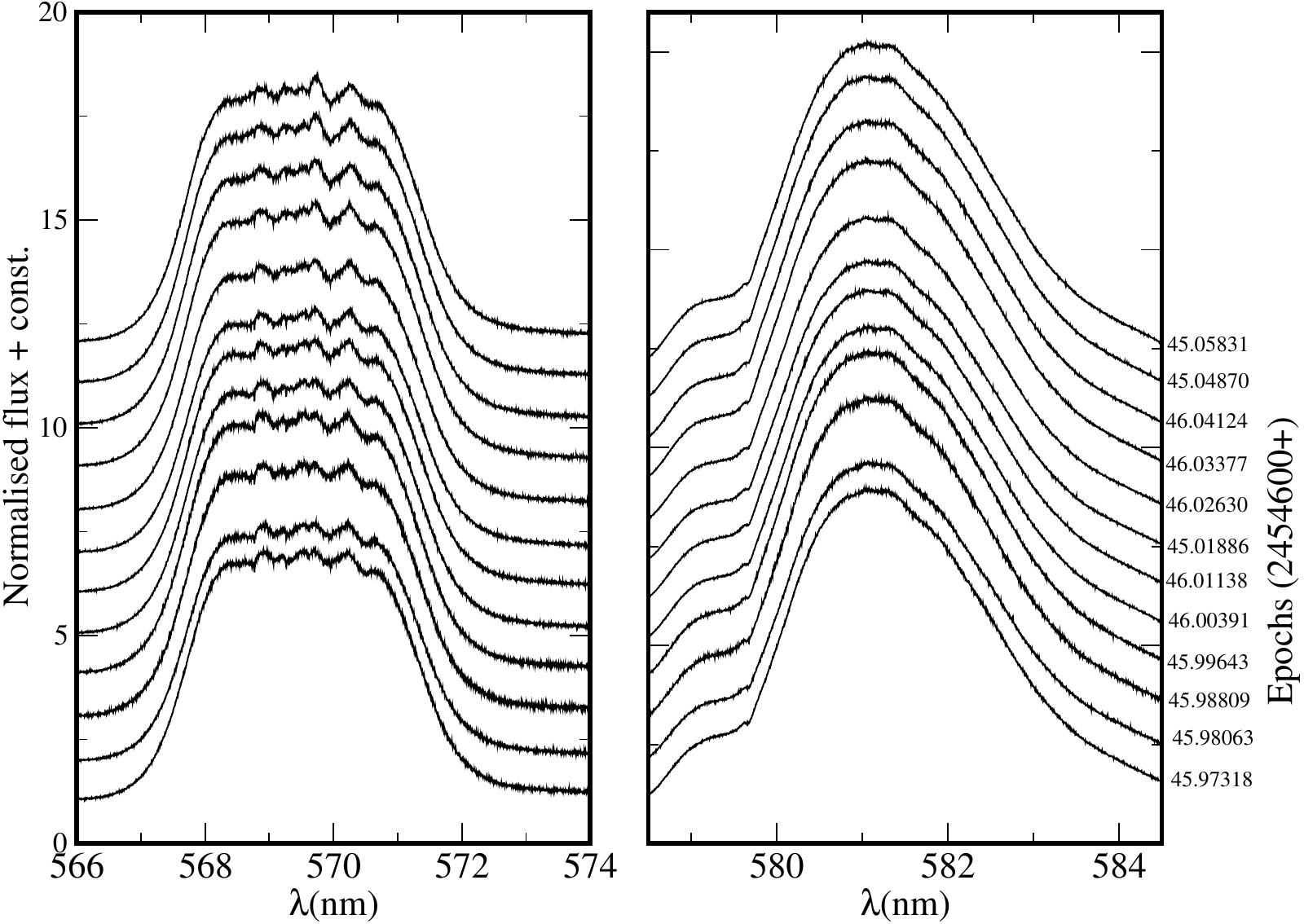}
    \caption{Line profile of C\,$\textsc{iii}$ $\lambda$5696 (left panel) and C\,$\textsc{iv}$ $\lambda\lambda$5802-12 (right panel) in the spectroscopic time series dataset from CFHT (see Table \ref{tab:observation}).}
    \label{fig:emission_lpv}
\end{figure}
\subsubsection{2m-Himalayan Chandra Telescope}\label{subsubsec:hct}
Low-resolution (R$\sim$1400$-$2200) spectra (SNR$\sim$40) of WR\,135 were observed using the HFOSC (Hanle Faint Object Spectrograph) mounted at the 2m-Himalayan \textit{Chandra} Telescope (HCT) situated at Hanle, Ladakh, India operated by Indian Institute of Astrophysics (IIA). Grism-7 (3800$-$7000\,\AA)\, and Grism-8 (5200$-$9000\,\AA)\, were used to cover the optical band. For wavelength calibration, Argon and Neon lamp spectra were recorded immediately following the capture of each science spectrum. Additionally, on the same night, we observed Feige 110 \citep{1990ApJ...358..344M}, a spectrophotometric standard star for the flux calibration, using Grisms 7 and 8. All the observations were recorded with a 2k$\times$4k CCD. The log of the observation is shown in Table \ref{tab:observation}.
We processed the 2D spectral images with IRAF \citep{1993ASPC...52..173T}, which included master bias subtraction, cosmic ray removal, wavelength and flux calibration of the 1D spectra extracted from the 2D spectral images. The reduction methods are discussed in detail in \citet{Kar_2024}.

\subsection{Data Analysis}\label{sect:analysis}
\subsubsection{Frequency analysis}
\begin{figure*}
\gridline{
    \fig{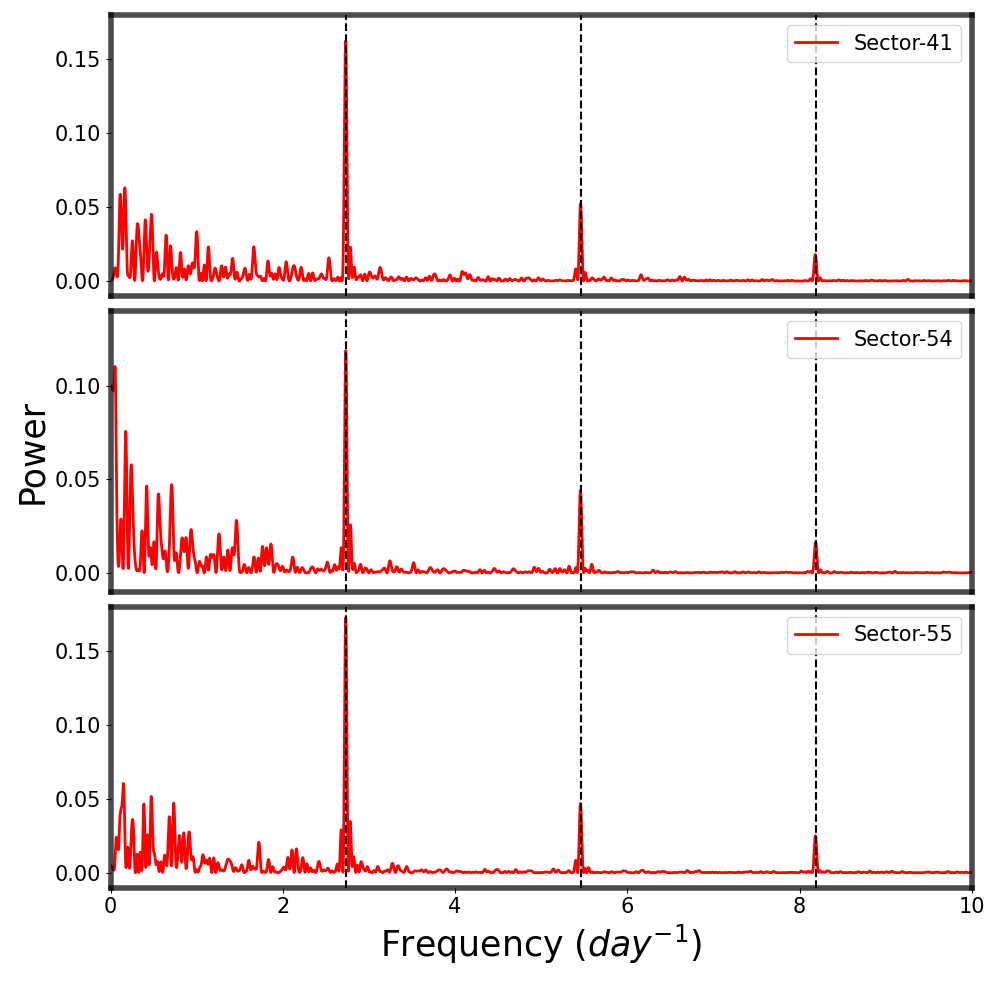}{0.5\textwidth}{(a) Harmonics from $1^{st}-3^{rd}$ order}
    \fig{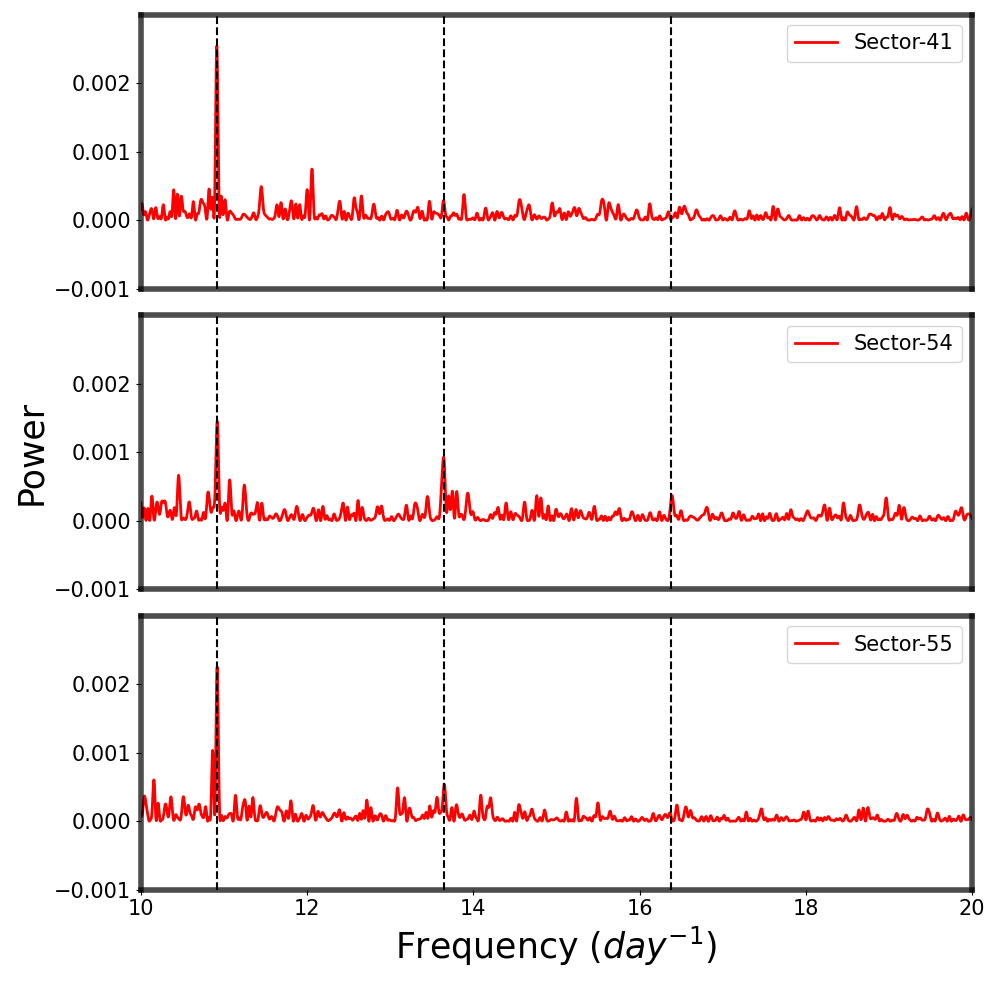}{0.5\textwidth}{(b) Harmonics from $4^{th}-6^{th}$ order}
    }
  \caption{Power spectra of WR\,135 generated from the LSP fitted photometric lightcurves from TESS Sector-41, 54 \& 55.}
  \label{fig:TESS_pow_spec}
\end{figure*}

\begin{figure*}
   \centering
   \includegraphics[width=0.8\textwidth, angle=0]{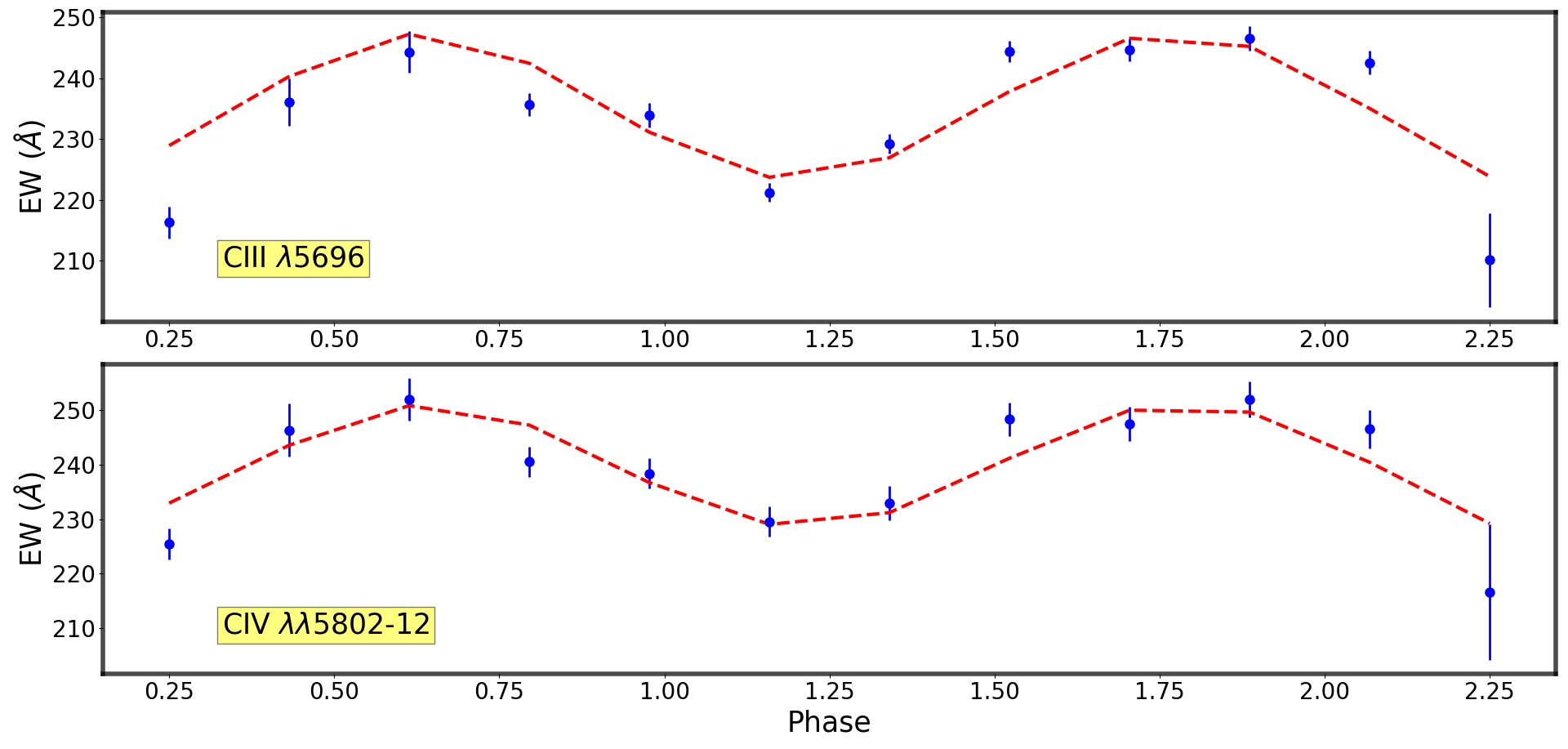}
   \caption{Dual phase sinusoids ($\nu$ = 21$\pm$1 $day^{-1}$) fitted over the observed line strength time series.}
   \label{fig:CFHT_lightcurve}
   \end{figure*}

To explore whether the source is variable, we perform Fourier transformation of the TESS lightcurves (see Fig. \ref{fig:TESS_pow_spec}) from Sectors-41, 54, and 55 as well as the EW time-series (see Fig.\ref{fig:CFHT_lightcurve}) of the emission lines of C\,$\textsc{iii}$ 5696 and C\,$\textsc{iv}$ 5802-12 using PERIOD04\,\citep{2005CoAst.146...53L}. Following \citet{10.1093/mnras/sty2743}, we extract the frequencies with SNR$\geq$3 (for detecting low-amplitude higher-order harmonics) by iteratively fitting co-sinusoids to the lightcurves. Also, following \citet{2022MNRAS.514.2269T}, we estimate the optimal uncertainties to the frequencies by applying a non-linear least-square fit to the chosen frequencies simultaneously.

We verify the estimated frequencies by applying Generalised Lomb-Scargle Periodograms\footnote{\url{https://pyastronomy.readthedocs.io/en/latest/pyTimingDoc/pyPeriodDoc/gls.html}} (Generalised LSP) derived by \citet{2009A&A...496..577Z}, to each of the lightcurves. The GLSP takes into account measurement errors and a constant term while fitting the trigonometric function. We apply the error-weighted Lomb-Scargle periodogram using the normalization method devised in \citet{2009A&A...496..577Z}.

\subsubsection{Spectral modeling}
To understand the physical and chemical properties of the object, we employ the 1D radiative transfer code CMFGEN (\citealt{1998ApJ...496..407H, 2003IAUS..212...70H, 2012IAUS..282..229H}) for modeling the spectra. This code solves the radiative transfer equations in a spherically expanding atmosphere of a co-moving frame that is kept under statistical and radiative equilibrium conditions. It uses a linearization technique to ensure consistency between temperature structure, radiation field, and atomic level populations. A converged model is generated iteratively as the radiation field and level populations are interdependent.

The stellar surface of the WR stars is buried within an optically thick region. Therefore, in all our models, we consider the hydrostatic stellar core to be inert as it has no significant effect on either the optical emission line strengths or the continuum. However, following the stellar evolutionary studies, we consider the stellar radius ($R_{\ast}$) at a Rosseland optical depth of 20 \citep{2019A&A...621A..92S} and estimate the effective temperature ($T_{\ast}$) using the Stefan-Boltzmann law. The temperature and radius at the photosphere are determined at an optical depth of $\tau_{ross}$=2/3 using the gray-body approximation.

For WC-type stars, the stellar wind velocity follows a two-component $\beta$ velocity law \citep{1989ApJ...347..392H, 2022ApJ...924...44A, Kar_2024} which has been inferred from the line widths of different ionization states of the same atomic species. The scale height ($\Delta$R) of the velocity law is used to achieve consistency between the velocities at the inner boundary and supersonic winds of the atmosphere. All the emission lines are treated under Sobolev approximation following the \citet{1975ApJ...195..157C} wind approach. Red-shifted wings in emission line profiles are attributed to wind velocities exceeding thermal ion velocities. CMFGEN incorporates clumping empirically using a velocity-dependent volume filling factor (VFF) relation. The cooling mechanism in the winds is accounted for using two phenomena -- radiative recombination in the inner winds, and free-free and thermal collisions in the outer winds.

The emission lines in WC-type stars are produced from different types of atomic transitions \citep{1989ApJ...347..392H, 2011Ap&SS.336...87H}. The C\,$\textsc{iv}$ $\lambda\lambda$5802-12 transition is a result of continuum fluorescence while C\,$\textsc{iii}$ $\lambda$5696 is a product of the dielectronic recombination process. Dense stellar winds undergo multiple line scattering (line-blanketing) which is treated using the super-level approach. Considering a He-burning stellar core with the atmosphere depleted from both hydrogen and nitrogen, we select atomic opacity data including collision and oscillator strengths, photoionization cross-sections, and dielectronic recombination rates for various ionization states of the abundant elements (He\,\textsc{i-ii}, C\,\textsc{ii-iv}, O\,\textsc{ii-iv}, Ne\,\textsc{ii-iv}, Si\,\textsc{iii-iv}, S\,\textsc{iii-vi}, Ar\,\textsc{iii-v}, Ca\,\textsc{ii-vi} and Fe\,\textsc{iii-ix}), to address the mechanisms related to the radiative driving of the stellar winds.

\section{Results}\label{sec:results}
\subsection{Detection of variability}\label{subsec:harmonics}
We detect several harmonics (in Table \ref{tab:frequencies}) of the fundamental frequency (2.729\,$day^{-1}$) from the Fourier transform (see Fig. \ref{fig:TESS_pow_spec}) of the photometric lightcurves (see Fig. \ref{fig:CFHT_lightcurve}). Among them, the 1st and 2nd harmonics were reported by \citet{2021MNRAS.502.5038N}, which were slightly deviated from those reported in \citet{2022ApJ...925...79L}. Among the detected Fourier components, the frequencies 2.729\,$day^{-1}$, 5.456\,$day^{-1}$ and 8.183\,$day^{-1}$ had SNR$>$7 while the $4^{th}$ harmonic with a period of $\sim$2.198\,hours had SNR$>$4 across all three Sectors. Although weak, we still note $5^{th}$ (P$\sim$1.758\,hours) and $6^{th}$ (P$\sim$1.465\,hours) harmonics (with SNR$>$3) only in Sector-54 probably due to detector sensitivity.
\begin{deluxetable*}{lcccccccc}
\tablecaption{Frequencies (SNR$\geq$3) detected across different time series observations.\label{tab:frequencies}}
\tablewidth{0pt}
\tablehead{
\colhead{Dataset} & \colhead{1st} & \colhead{2nd} & \colhead{3rd} & \colhead{4th} & \colhead{5th} & \colhead{6th} & \colhead{7th} & \colhead{8th}\\
\colhead{} & \colhead{($d^{-1}$)} & \colhead{($d^{-1}$)} & \colhead{($d^{-1}$)} & \colhead{($d^{-1}$)} & \colhead{($d^{-1}$)} & \colhead{($d^{-1}$)} & \colhead{($d^{-1}$)} & \colhead{($d^{-1}$)}
}
\startdata
 & & & Photometric (TESS) & & & & & \\ \hline
Sector-41 & 2.729$\pm$0.001 & 5.457$\pm$0.001 & 8.183$\pm$0.002 & 10.912$\pm$0.006 & - & - & - & - \\
Sector-54 & 2.730$\pm$0.001 & 5.456$\pm$0.001 & 8.186$\pm$0.002 & 10.915$\pm$0.008 & 13.642$\pm$0.009 & 16.386$\pm$0.015 & - & - \\
Sector-55 & 2.730$\pm$0.001 & 5.456$\pm$0.001 & 8.185$\pm$0.001 & 10.915$\pm$0.005 & 13.65$\pm$0.012 & - & - & - \\ \hline
 & & & Spectroscopic (CFHT) & & & & & \\ \hline
$\mathrm{EW_{C\,\textsc{iii} \lambda5696}}$ (\AA) & - & - & - & - & - & - & - & 21$\pm$1 \\
$\mathrm{EW_{C\,\textsc{iv} \lambda\lambda5802-12}}$ (\AA) & - & - & - & - & - & - & - & 21$\pm$1 \\
\enddata
\end{deluxetable*}  

We do not re-establish the red and white-noise level of the data as those have been already derived in the earlier studies by fitting a semi-Lorentzian model to the light curves. \citet{2022ApJ...925...79L} found that the mean amplitude of the frequency-dependent red-noise for a 10\,minutes cadence is almost similar to that for a 30\,minutes cadence data while the characteristic frequency remained same in each case. This was drawn from the fact that the mean of the frequency-dependent red-noise amplitude for the 10\,minutes, and 30\,minutes cadence data was found to lie within one standard deviation of the red noise of the 2\,minutes cadence. The red-noise (1.512 $day^{-1}$) detected earlier \citep{2022ApJ...925...79L, 2021MNRAS.502.5038N} in WR\,135 has been attributed to stochastic variability. The pulsation fundamental frequency is almost twice that of the stochastic variability frequency indicating that the wave propagation velocity is higher than the dynamical outflow velocity.

Among the emission lines, C\,$\textsc{iii}$ $\lambda$5696; C\,$\textsc{iv}$ $\lambda$5802-12; and He\,$\textsc{i}$ $\lambda$5876 are found to be the least blended with their adjacent lines. The profiles corresponding to these emission lines are shown in Fig. \ref{fig:emission_lpv}. From the LSP analysis of the time series of measured EWs (see Fig. \ref{fig:CFHT_lightcurve}), we find that the first two of them show a sinusoidal variability of frequency $\sim$22 $day^{-1}$. It must be noted that the detected frequency is the 8$^{th}$ harmonic of the fundamental frequency (2.729 $day^{-1}$).

We do not detect epoch-dependant frequencies to what is reported in WR\,7 \citep{2022MNRAS.514.2269T} which was found to exhibit not only CIR-type oscillations but also harmonics of its fundamental pulsation. Therefore the detected oscillations cannot be due to CIRs as they are persistent across all three epochs. 

\subsection{Modeled spectra}\label{subsec:model}
Using CMFGEN, we derive the atmospheric models for the observed Spectral energy Distribution (hereafter, SED) by varying the physical and chemical parameters (see Table \ref{tab:model}). Tailored spectroscopic models are generated to consistently reproduce the emission lines and the continuum of the SED across optical to IR wavebands. While fitting with the observed data, the models were adjusted for the spectrograph resolution. The upper and lower bounds to the physical (i.e. luminosity, radius, mass loss rate, etc.) and chemical (i.e. carbon/helium abundance) parameters are derived from the relative line strengths (C\,\textsc{iv}/C\,\textsc{iii}; C\,\textsc{iv}/He\,\textsc{i}) of the diagnostic emission lines for WCL-type stars, as mentioned in \citet{Kar_2024}.
\begin{deluxetable}{ccc}
\tablecaption{Best spectroscopic model fitting parameters \label{tab:model}}
\tablewidth{0pt}
\tablehead{
\colhead{Parameters} & \colhead{Values} & \colhead{}
}
\startdata
$\log L_{\ast} (L_{\odot})$  & $5.72_{-0.15}^{+0.15}$ & \\
$T_{\ast}$ (K) & 60740 & \\
$R_{\ast} (R_{\odot})$ & $6.56_{-0.5}^{+0.5}$ & \\
$T_{2/3}$ (K) & 53400 & \\
$R_{2/3} (R_{\odot})$ & 8.49 & \\
$\log \dot{M} (M_{\odot}/yr)$ & $-4.66_{-0.032}^{+0.035}$ & \\
f & 0.29 & \\
$\log \dot{M}/\sqrt{f} (M_{\odot}/yr)$ & -4.167 & \\
$v_{core} (km s^{-1})$ & 0.4 & \\
$v_{phot} (km s^{-1})$ & 100 & \\
$v_{dop} (km s^{-1})$ & 50 & \\
$v_{\infty,1} (km s^{-1})$ & 1000 & \\
$v_{\infty,2} (km s^{-1})$ & 1400 & \\
$\beta_{1}$ & 1 & \\
$\beta_{2}$ & 50 & \\
$\eta$ & 2.623 & \\
$M (M_{\odot})$ & 20.89 & \\ \hline
{Atomic Elements} & \colhead{Relative fraction} & \colhead{Mass fraction} \\ \hline
H & 0.00 & 0.00 \\
He & 1.00 & 0.542 \\
C & $0.25_{-0.24}^{+0.15}$ & $0.406_{-0.374}^{+0.111}$ \\
N & 0.00 & 0.00 \\
O & 0.02 & 0.043 \\
Ne & $2.2e^{-3}$ & $6.02e^{-3}$ \\
Si & $2.26e^{-4}$ & $8.62e^{-4}$ \\
S & $8.75e^{-5}$ & $3.82e^{-4}$ \\
Ar & $1.92e^{-5}$ & $1.04e^{-4}$ \\
Ca & $1.12e^{-5}$ & $6.15e^{-5}$ \\
Fe & $2.18e^{-4}$ & $1.66e^{-3}$ \\ \hline
$E_{B-V}$ & 0.5 & \\
$A_{V}$ & 1.55 & \\
\enddata
\end{deluxetable}

\begin{figure*}
  \gridline{
    \fig{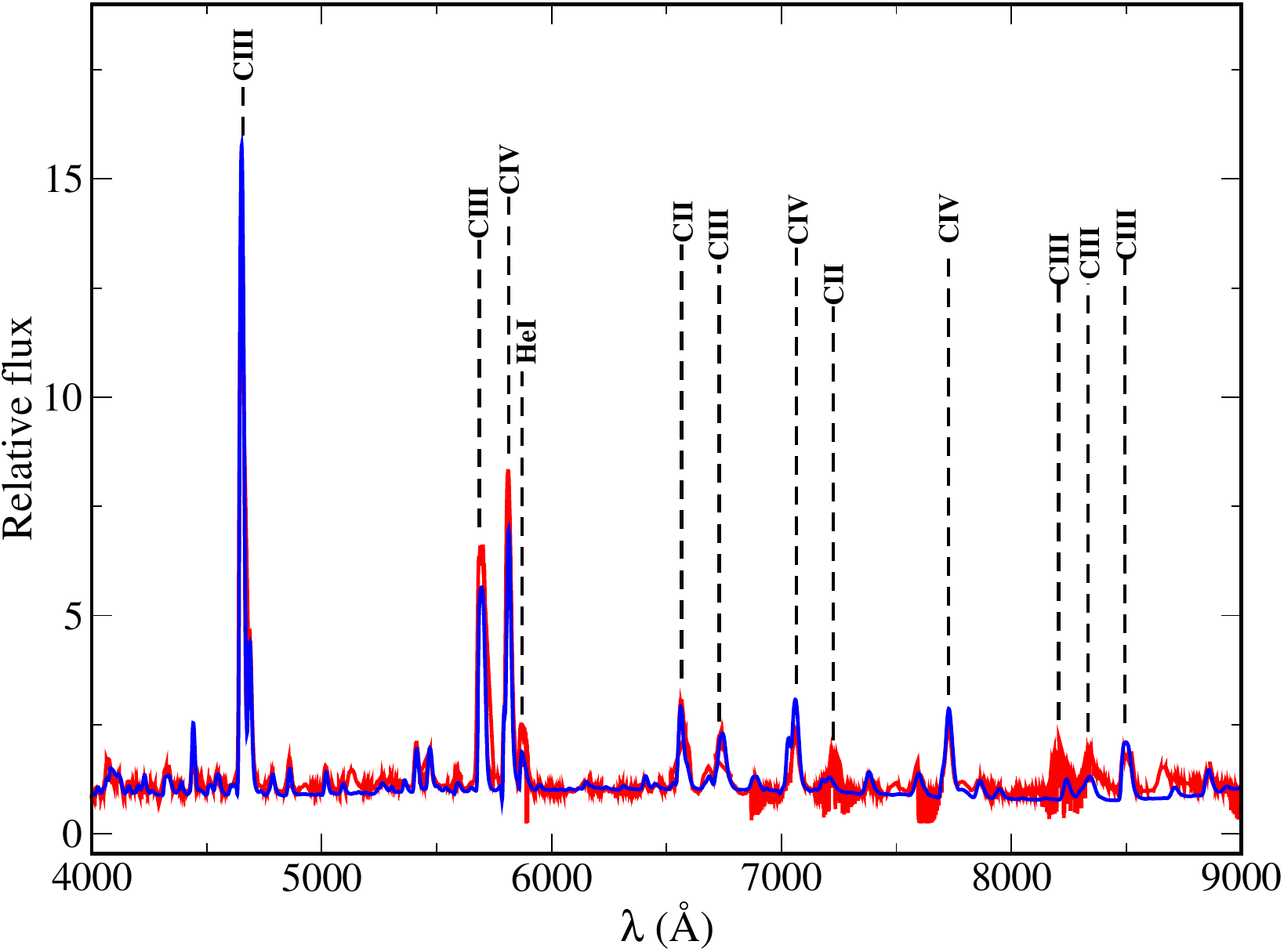}{0.6\linewidth}{(a) Normalized observed spectrum (represented in \textit{red}) of WR\,135 fitted with continuum normalized model (represented in \textit{blue}) spectrum.}
    }
  \gridline{
    \fig{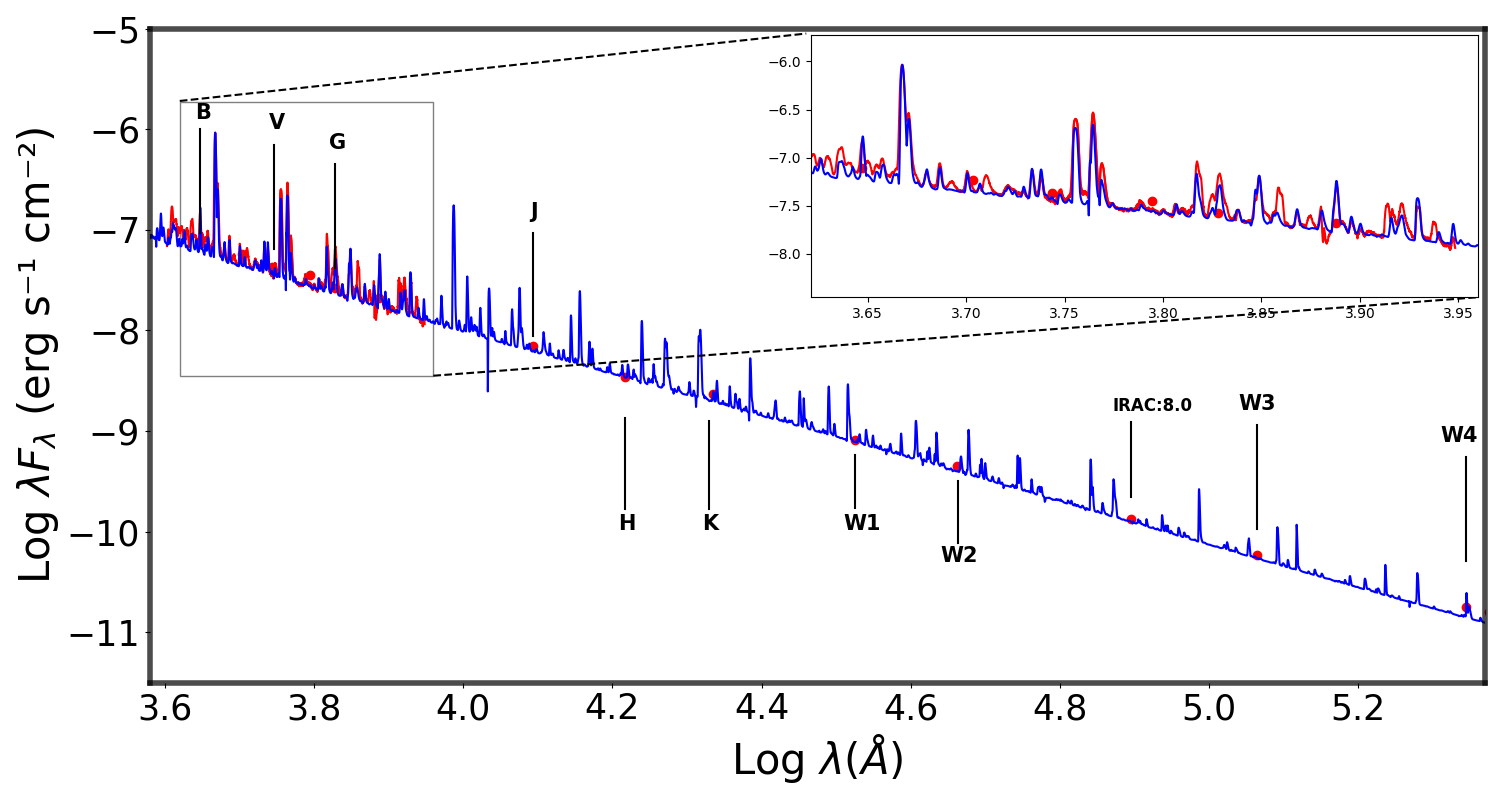}{0.8\textwidth}{(b) De-reddened observed SED (in \textit{red}) fitted with distance-scaled model SED (in \textit{blue}) across 4000-221000\,\AA.}
    }
  \caption{Comparison of the best fitting spectroscopic model with the observed data of WR\,135.}
  \label{fig:SED_fitting}
\end{figure*}

In Fig.\ref{fig:SED_fitting} (a), we show that our model consistently predicts the line strengths of most of the emission lines including the targeted ones i.e.  C\,$\textsc{iii}$ $\lambda$5696 and C\,$\textsc{iv}$ $\lambda\lambda$\,5802-12 observed in the normalized optical spectrum of WR\,135. From the estimated model parameters (see Table \ref{tab:model}), we note that our model has a larger core radius, and higher luminosity but similar temperature and carbon abundance compared to that estimated by \citet{2019A&A...621A..92S} for WR\,135. The chemical composition for the rest of the elements is chosen from earlier studies \citep{2012A&A...540A.144S} of WC8-type objects. The mass of the star ($\sim$21\,$M_{\odot}$) is derived using the mass-luminosity relationship \citep{2011A&A...535A..56G} for He-burning chemically homogenous WC-type stars. We de-redden and fit (see Fig.\ref{fig:SED_fitting} (b)) the observed SED using the average Interstellar extinction law: G23 (\citet{2009ApJ...705.1320G,2019ApJ...886..108F, 2021ApJ...916...33G, 2022ApJ...930...15D, 2023ApJ...950...86G}, distributed as $\textit{dust\_extinction}$ by \textit{Astropy}) and scale the model SED using the latest GAIA DR2 \citep{2018AJ....156...58B, 2018A&A...616A...1G} distance 2.36\,kpc \citep{10.1093/mnras/stz3614}. From the estimated physical parameters, we see that the choice of the velocity law and clumping factor play a major role in fitting the emission lines of different species. In the next subsections (Sec. \ref{subsec:vel_grad} and Sec. \ref{subsec:clump}) we discuss the influence of wind acceleration and size of the clumps present in the stellar atmosphere. 

\subsubsection{Impact of wind acceleration}\label{subsec:vel_grad}
\begin{figure*}[ht]
  \gridline{
    \fig{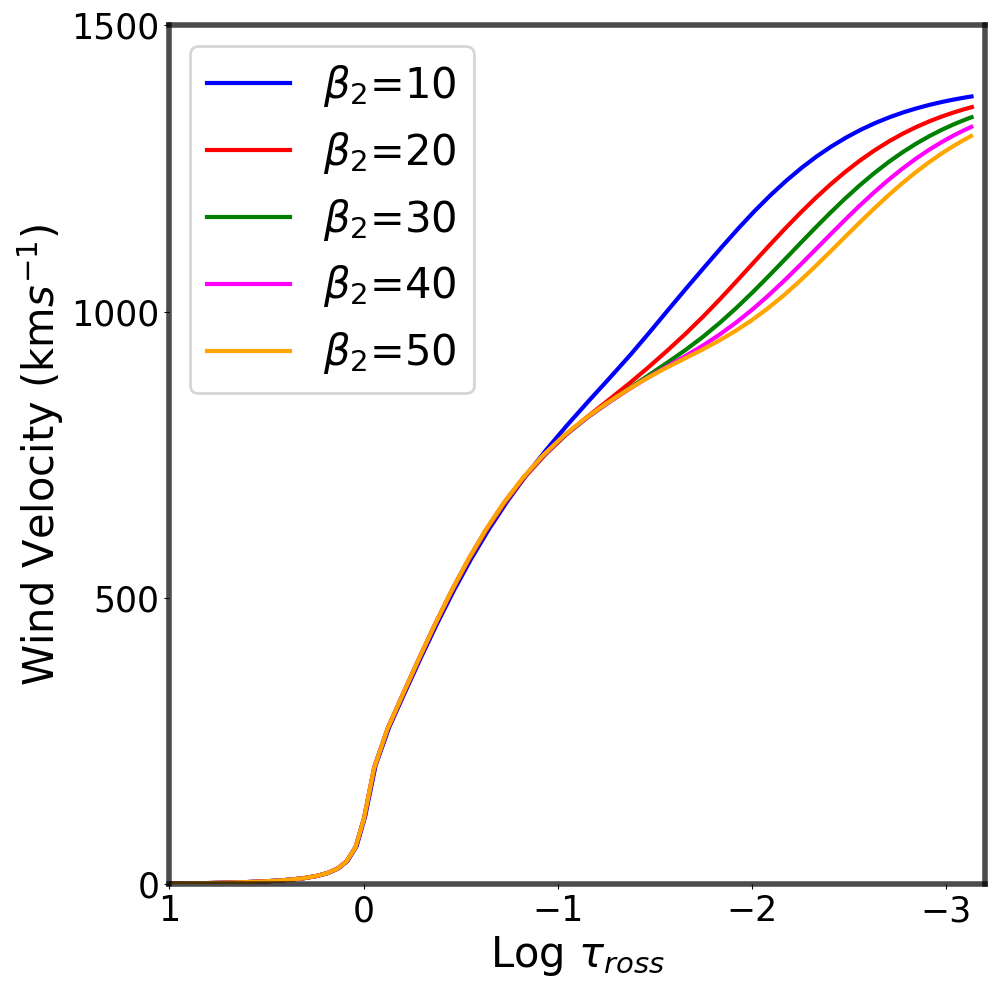}{0.45\linewidth}{(a) Velocity profiles corresponding to different $\beta_{2}$ values.}
    \fig{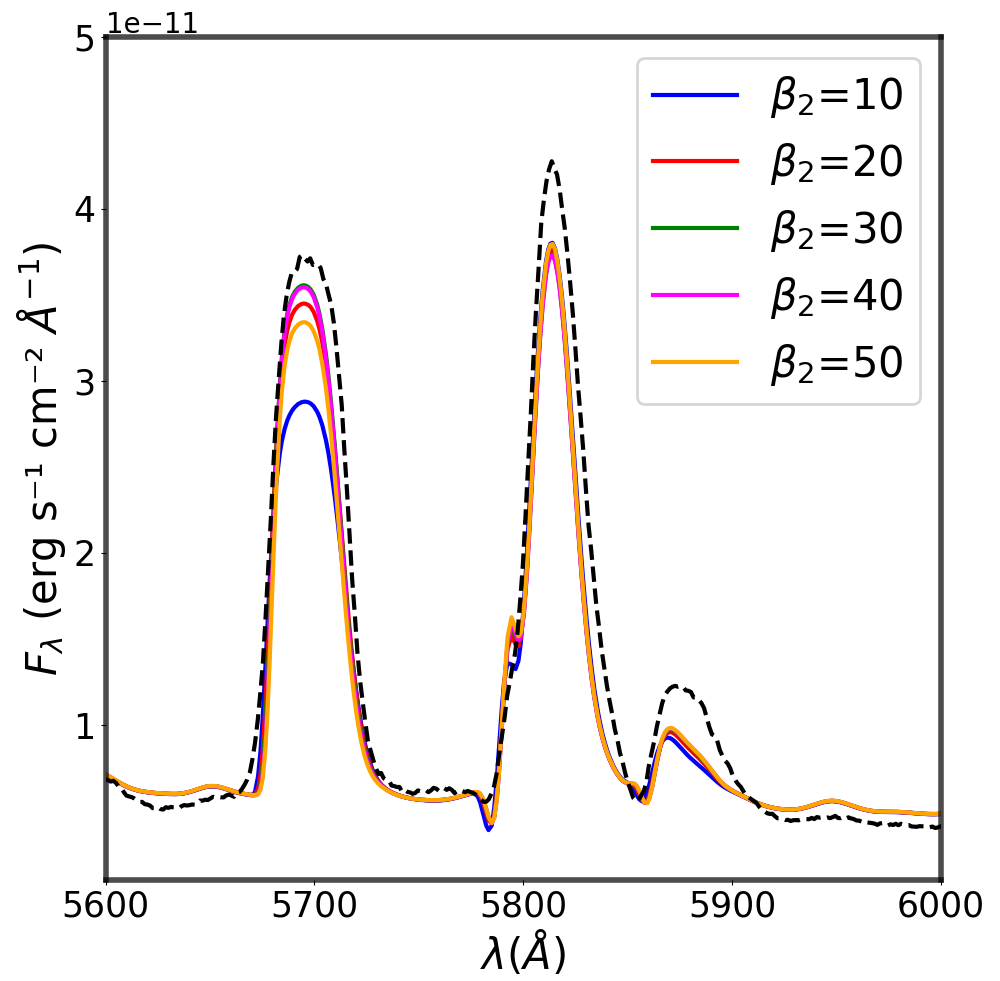}{0.45\linewidth}{(b) Variation in the emission line strengths. The black-dashed curve denotes the observed emission lines of the spectrum.}}
  
  \caption{Influence of acceleration exponent ($\beta_{2}$) on the unblended diagnostic emission lines (for $f_{VFF}=0.29$). The $\beta_{2}$-exponent affects the outer wind ($\tau_{ross}>$0.1) density which in turn affects the line strengths. }
  \label{fig:beta_compare}
\end{figure*}

The velocity gradient affects the mass-loss rate which in turn affects the emission line strengths. We observe the role of acceleration in the outer optically thin supersonic winds. Keeping other parameters fixed, we only vary the outer-wind acceleration exponent ($\beta_{2}$) and generate corresponding spectroscopic models (see Fig. \ref{fig:beta_compare}). 
We find that the line strength of the C\,$\textsc{iv}$ $\lambda\lambda$\,5802-12 remains almost unaffected while the C\,$\textsc{iii}$ $\lambda$5696 changes significantly with the choice of the outer acceleration exponent. Such results indicate that a significant number of C\,$\textsc{iii}$ ions undergo transitions in the outer winds. It is surprising to note that, C\,$\textsc{iii}$ $\lambda$5696 shows decreased line strengths for $\beta_{2}$=10 and 50; while the strength increases for $\beta_{2}$=20 to 40. 
Hence, to simultaneously fit both the emission lines, small-scale inhomogeneities should exist at the corresponding overlapped line formation region which can only be explained if a large number of small clumps are present in those winds.
\subsubsection{Clumping in the winds} \label{subsec:clump}
\begin{figure*}
  \gridline{
    \fig{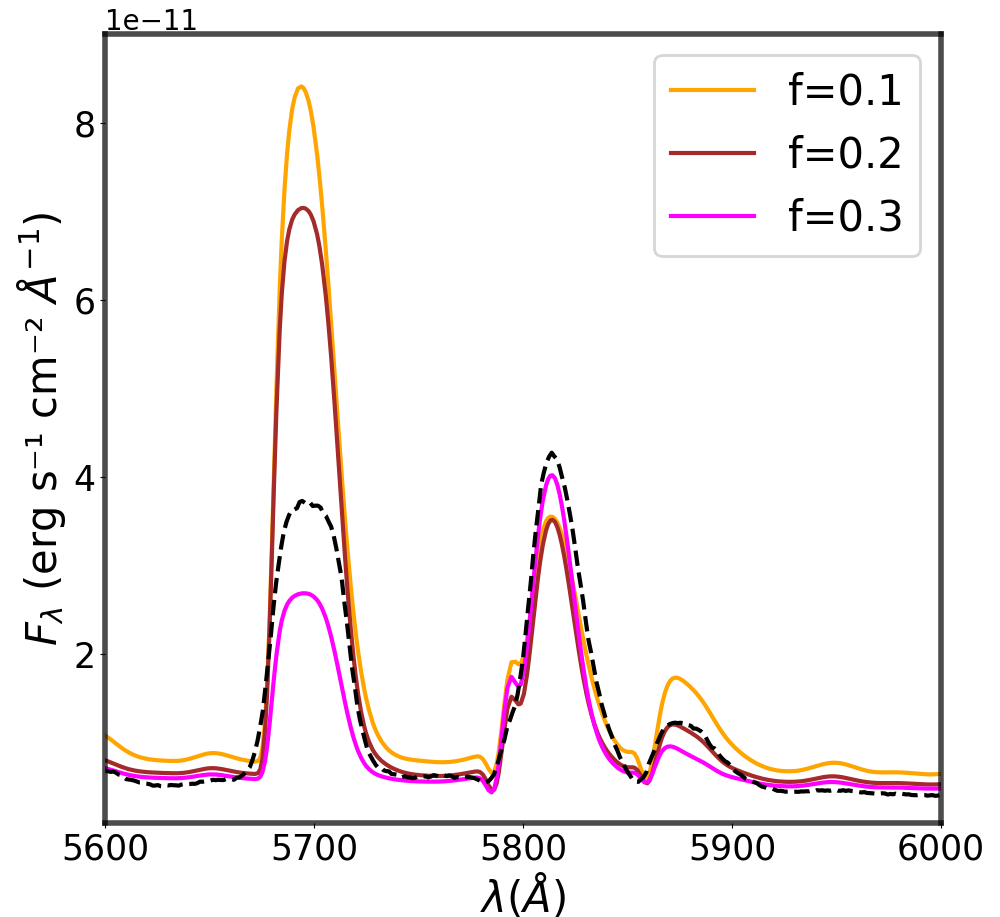}{0.5\linewidth}{(a) Clump size variation (coarse)}
    \fig{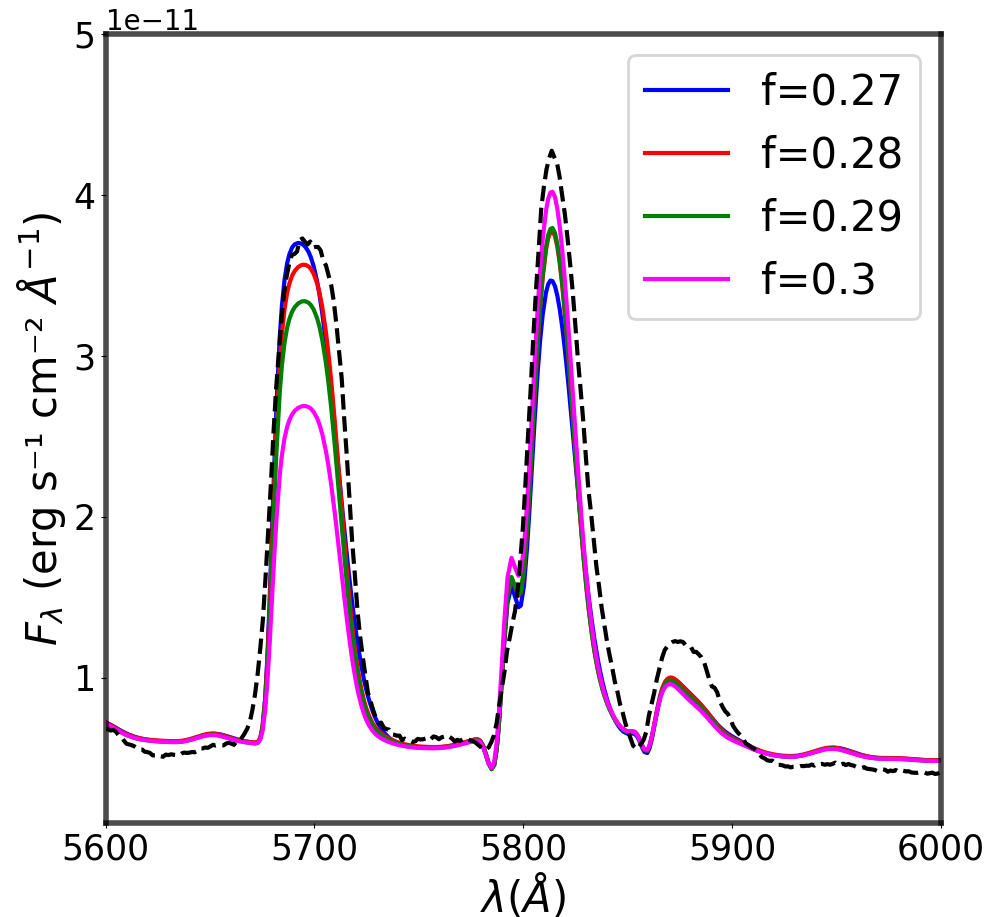}{0.5\linewidth}{(b) Clump size variation (fine)}
  }
  \caption{Influence of VFF on the unblended diagnostic emission lines for a fixed outer-wind acceleration exponent ($\beta_{2}=50$). The observed spectrum is represented as a black-dashed curve. Models computed for different VFF values are also labeled.}
  \label{fig:clump_compare}
\end{figure*}

We know that the recombination line strengths scale with density ($\rho$) and volume filling factor (VFF) as $\rho^{2}/f_{VFF}$. We vary the VFF (i.e. reciprocal of the clumping factor, $D_{cl}$ = 1/$f_{VFF}$), $f_{VFF}$ from $0.1-0.3$, and note their impact on the emission lines (see Fig. \ref{fig:clump_compare}). We note that bigger clumps ($f_{VFF}=0.1-0.2$) affect not only the line strengths but also the slope of the SED, while the smaller clumps affect only the line strengths. From Fig. \ref{fig:clump_compare}, we find that model and observed strengths of C\,$\textsc{iii}$ $\lambda$5696 and C\,$\textsc{iv}$ $\lambda\lambda$\,5802-12 fit best for $f_{VFF}=0.27-0.3$, while that for He\,$\textsc{i}$\,$\lambda$5876 at $f_{VFF}=0.2$. 
We also note that the C\,$\textsc{iii}$ $\lambda$5696 transition is more susceptible to the fractional change in the size of the clumps than C\,$\textsc{iv}$ $\lambda\lambda$\,5802-12. A slight change in the clumping size alters the line strengths drastically. This is possible if clumps of different sizes are present in the optically thin regions of the stellar winds. These are produced by the velocity perturbation due to the instabilities present at the base of the winds.

\subsection{Driving source}\label{subsec:driving}
We examine the physical phenomena that might be responsible for the launching of the stellar winds as well as the pulsations detected from the photometric and spectroscopic time series observations. The radiating envelopes of such an expanding stellar atmosphere are mainly driven by pressure and opacity. Therefore, we investigate their roles in the following subsections.
\subsubsection{Role of Pressure}
\begin{figure*}
  \gridline{
    \fig{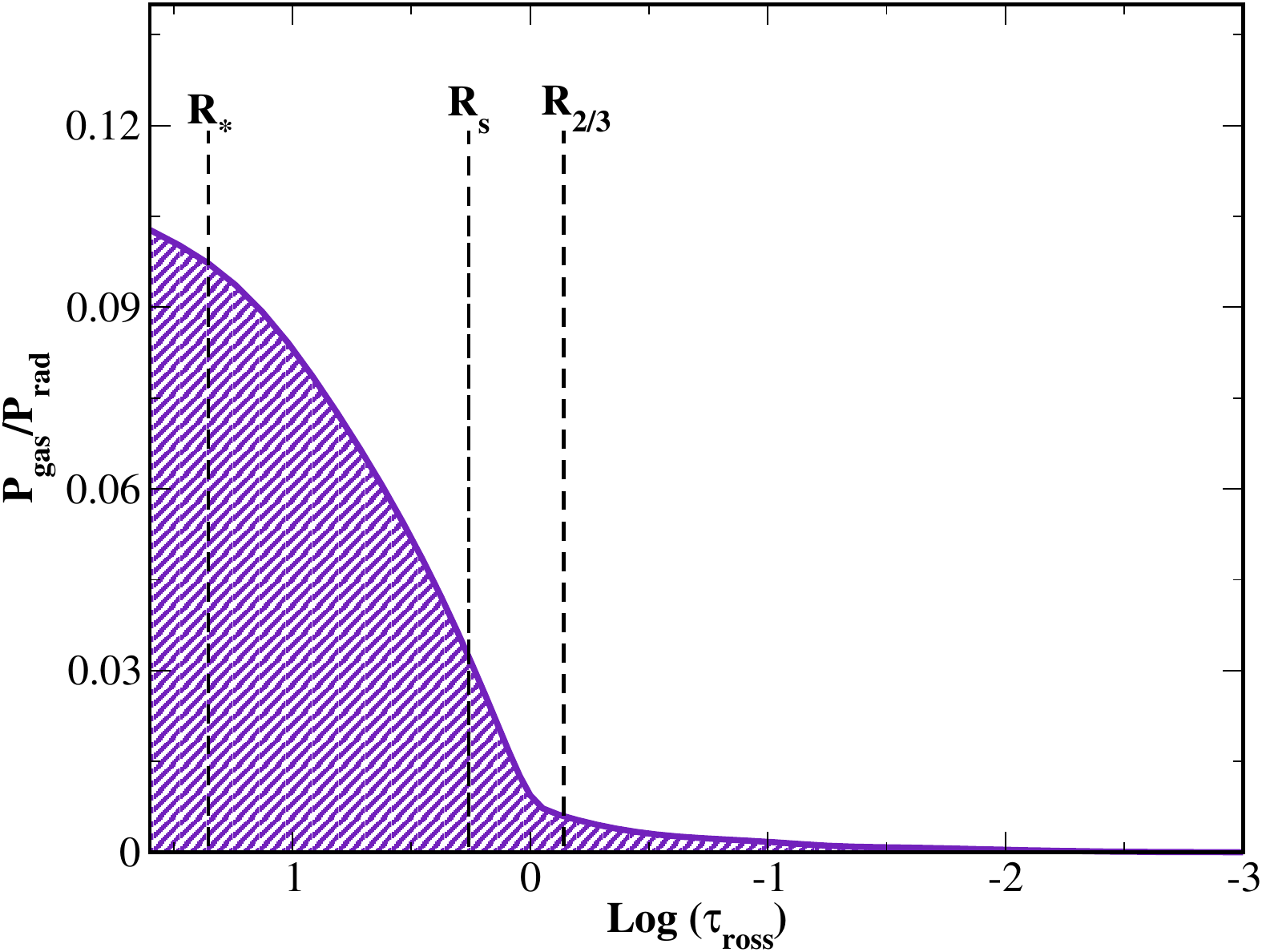}{0.5\linewidth}{(a) Overall}
    \fig{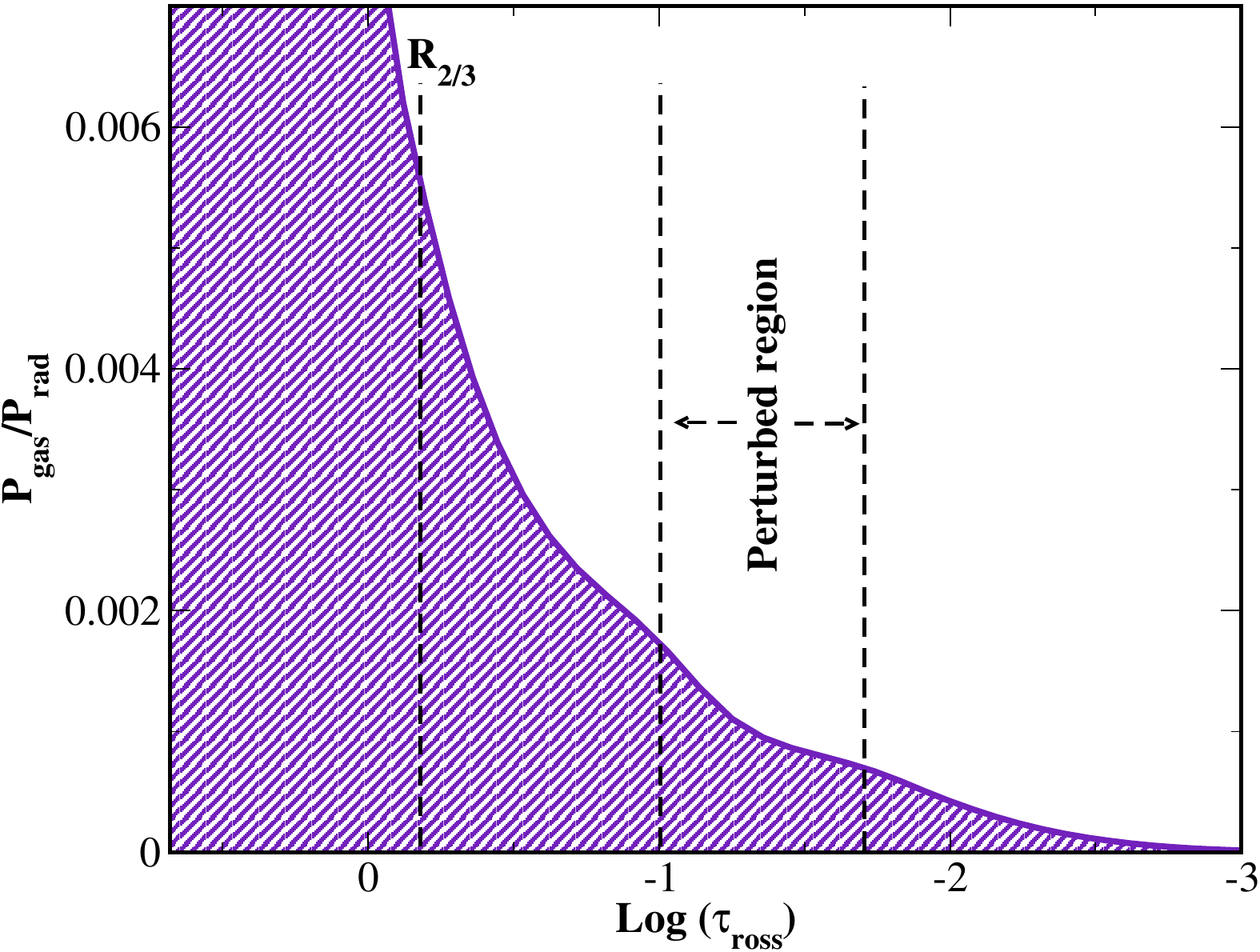}{0.5\linewidth}{(b) Wind}
  }
  \caption{Ratio of the gas pressure ($P_{\mathrm{gas}}$) and radiation pressure ($P_{\mathrm{rad}}$) in the stellar model atmosphere. In the supersonic regime, the $P_{\mathrm{gas}}$ becomes almost negligible compared to the $P_{\mathrm{rad}}$.}
  \label{fig:Pgas_Prad}
\end{figure*}

\begin{figure}
  \gridline{
    \fig{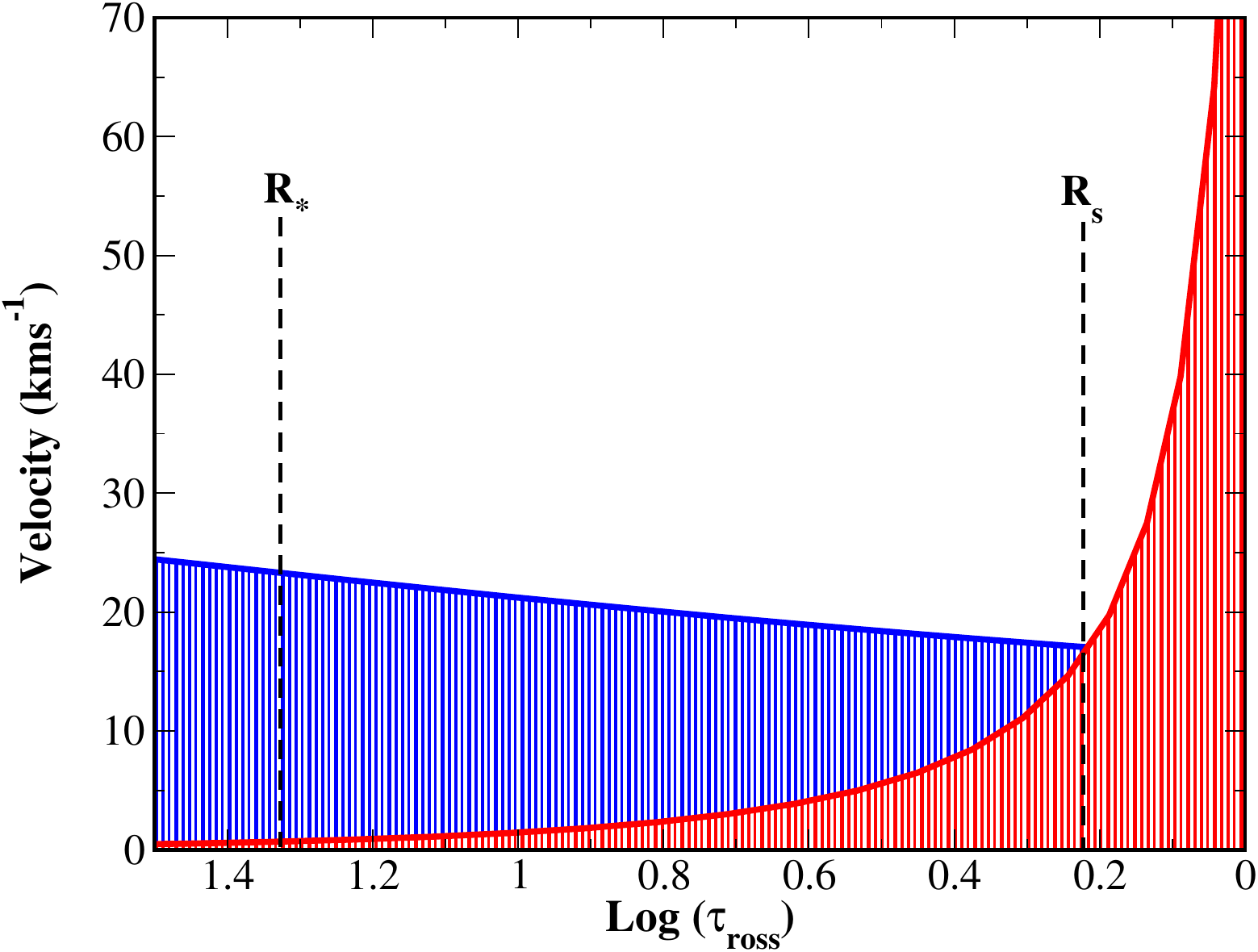}{0.5\linewidth}{(a) Inner layers}
    \fig{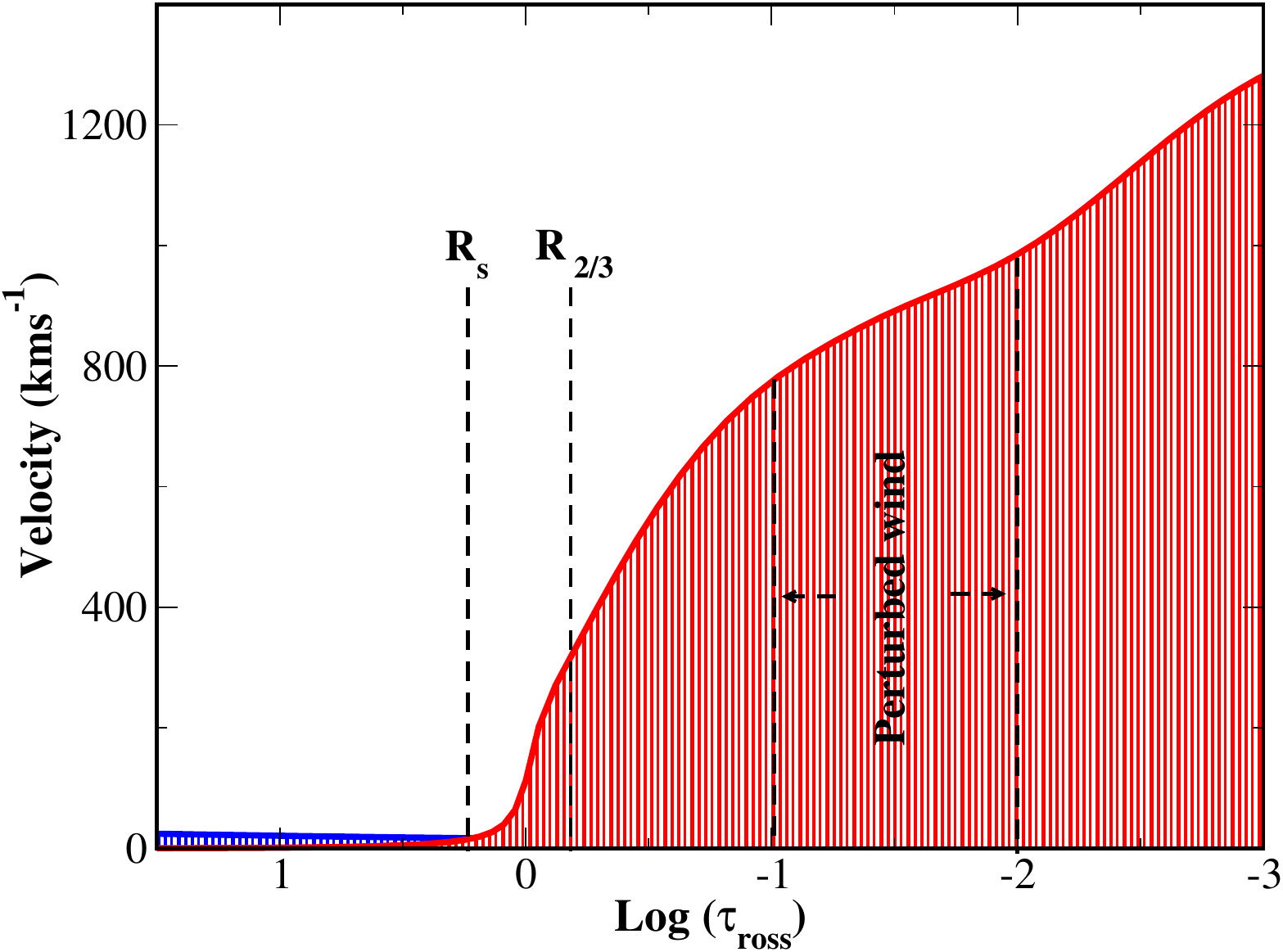}{0.5\linewidth}{(b) Outer layers}
  }
  \caption{Velocity stratification in the model atmosphere of WR135: Isothermal sound speed (in \textit{blue}) and Wind velocity (in \textit{red}). The mass-outflow follows the wind velocity profile beyond the sonic point $(R_{s})$.}
  \label{fig:vel_compare}
\end{figure}

From Fig.\ref{fig:Pgas_Prad} (a), we find a steady fall in the $P_{gas}/P_{rad}$ curve as the radiation pressure ($P_{rad}$) dominates the stellar atmosphere that lies beyond the stellar surface ($R_{\ast}$). However, above the photosphere ($\tau_{ross}\leq$1) where the wind reaches supersonic velocities, the slope of the curve becomes less steep. The isothermal sound speed ($\sim$25\,$kms^{-1}$) of the gas remains dominant up to the sonic point, $R_{s}$ (at $\tau_{ross}\sim1.7$). Beyond $R_{s}$ the stellar winds are more strongly driven by the radiation pressure (see Fig. \ref{fig:vel_compare} (a)). In the $P_{gas}/P_{rad}$ curve (see Fig. \ref{fig:Pgas_Prad} (b)), we note a slope change (marked as perturbed region) in the radiation pressure in the same optical depths ($\tau_{ross}=0.1-0.01$) as the velocity (see Fig. \ref{fig:vel_compare} (b)). This is caused by the two-component velocity law which resembles the nature of hydrodynamic wind detected by \citet{2020MNRAS.491.4406S} in the case of WR 111 (WC5).

\subsubsection{Impact of Opacity}
\begin{figure}
   \centering
   \includegraphics[width=0.7\textwidth, angle=0]{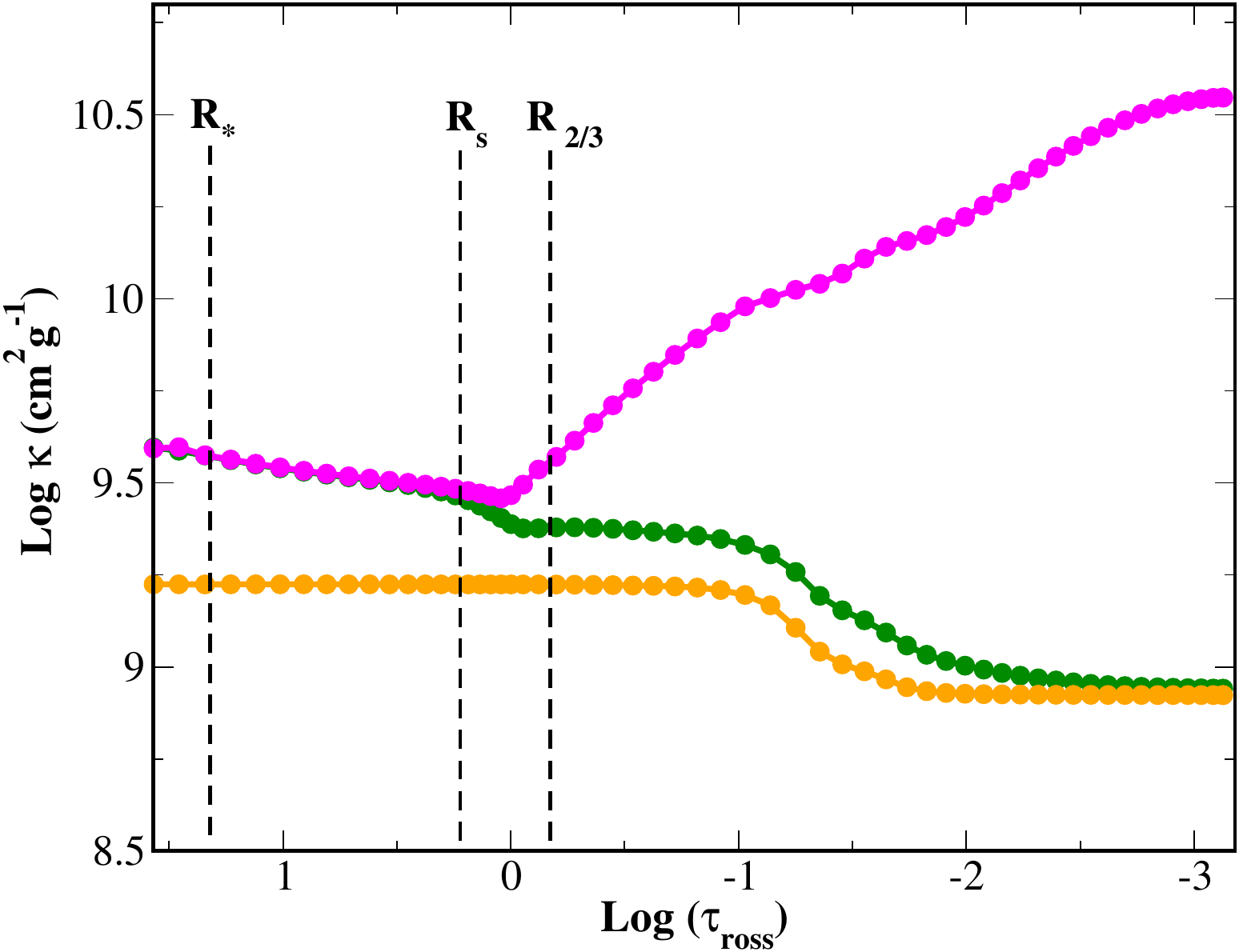}
   \caption{Comparison among the mean-flux (in \textit{magenta}), mean-Rosseland (in \textit{green}), and electron-scattering (or Thomson in \textit{orange}) opacities being contributed by different atomic species (in Fig. \ref{fig:ion_frac}) in the stratified winds of the model atmosphere.}
   \label{fig:opac_compare}
\end{figure}

\begin{figure*}
  \gridline{
    \fig{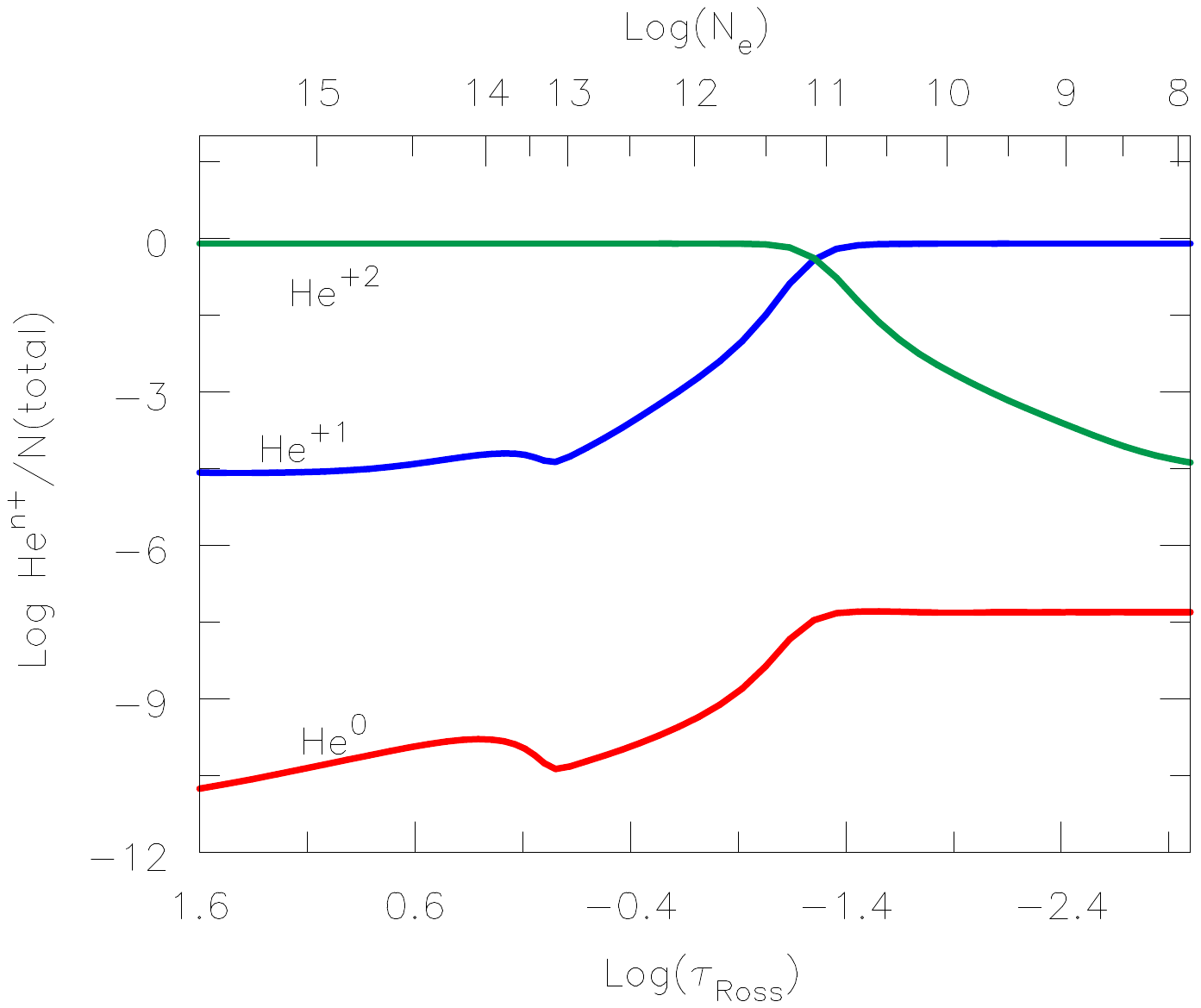}{0.5\linewidth}{(a) Helium}
    \fig{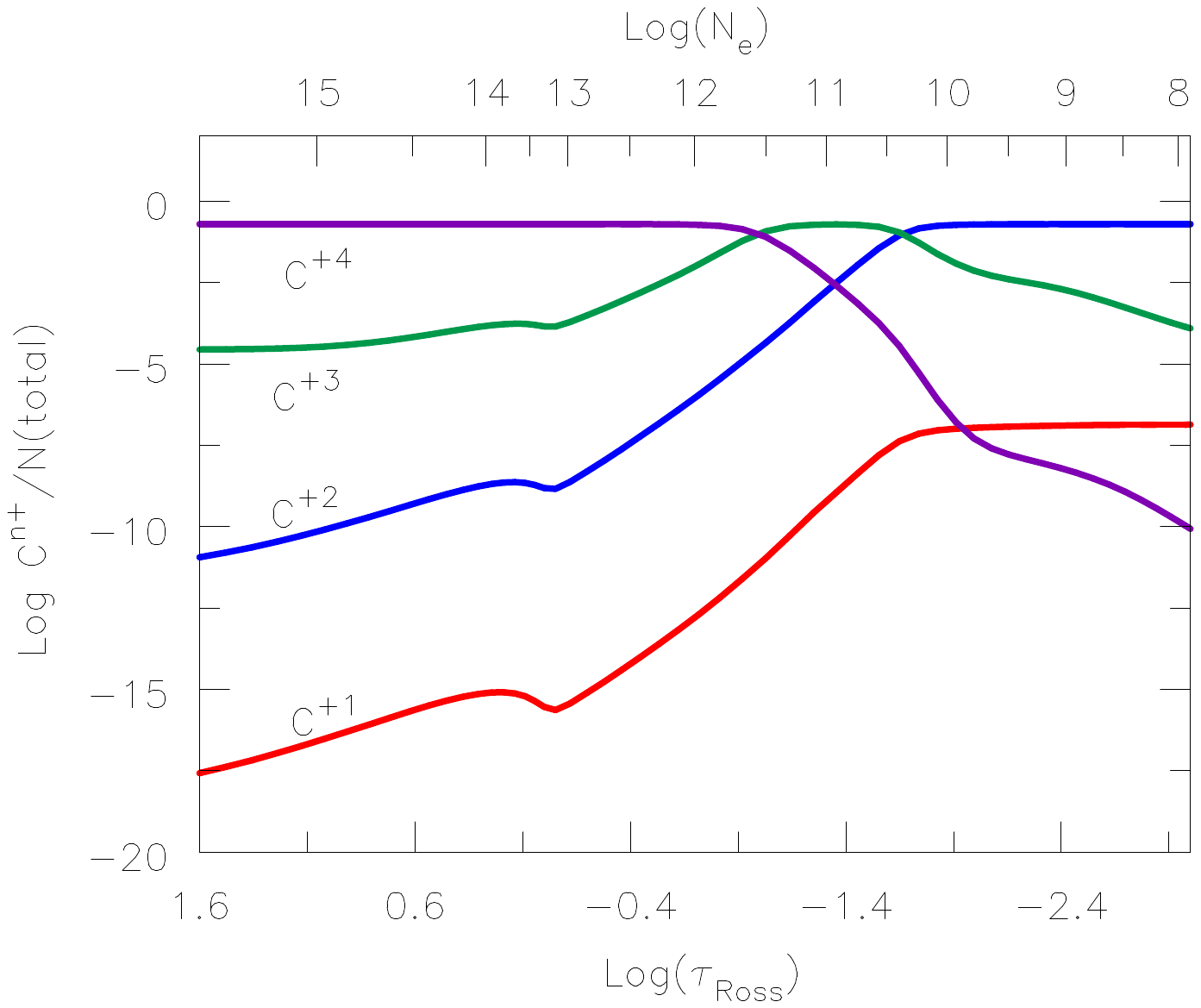}{0.5\linewidth}{(b) Carbon}
    }
  \gridline{
    \fig{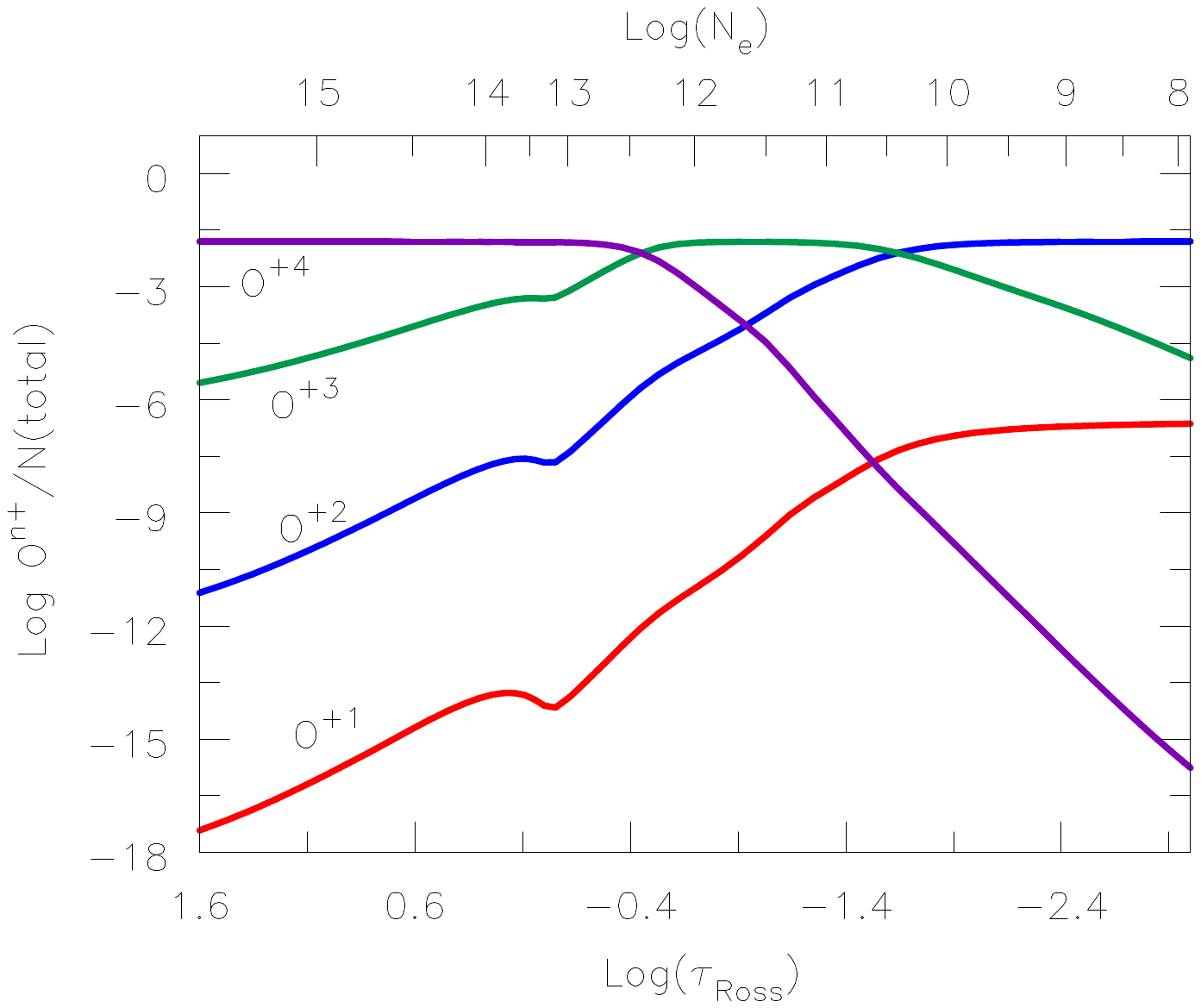}{0.5\linewidth}{(c) Oxygen}
    \fig{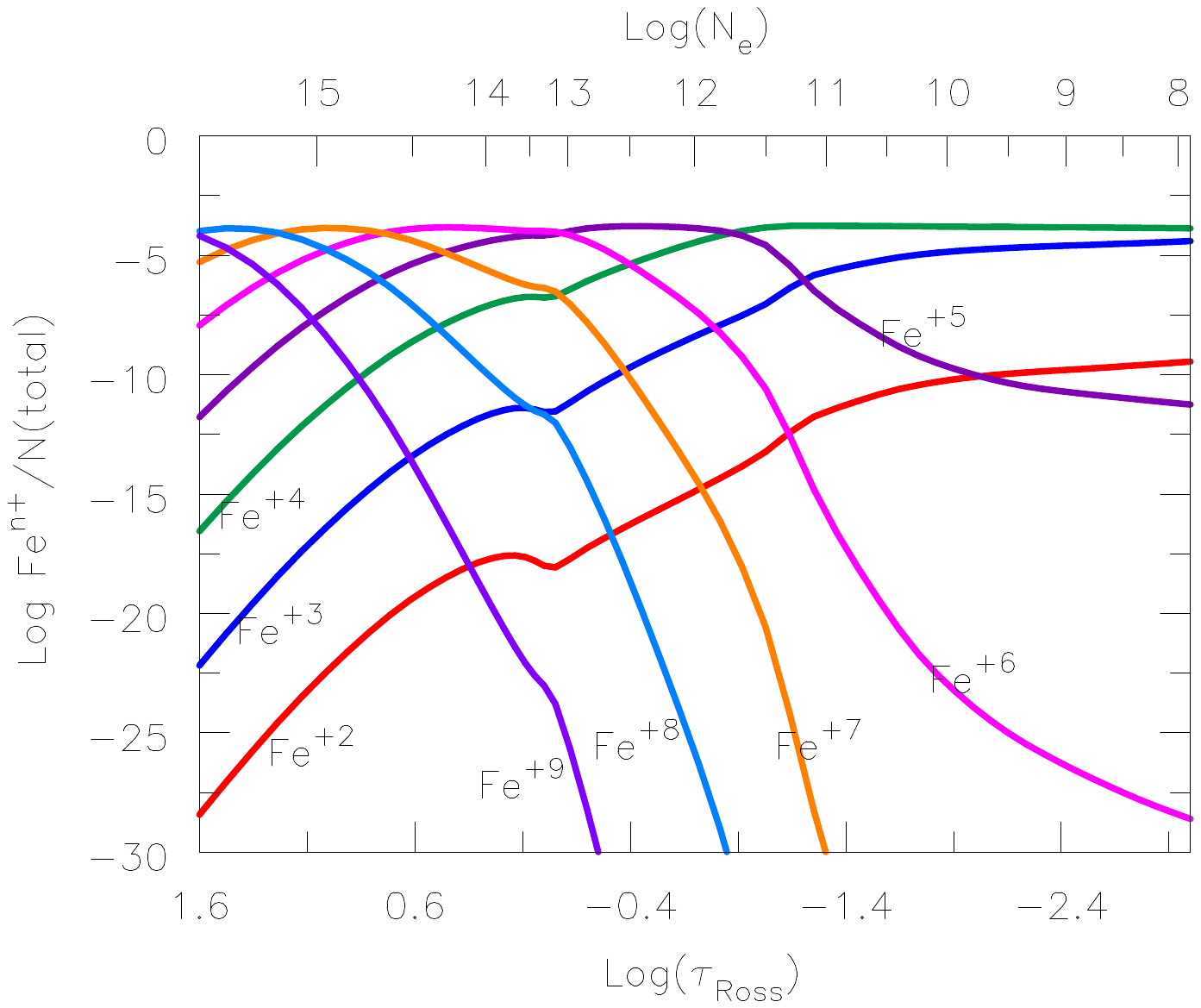}{0.5\linewidth}{(d) Iron}
  }
  \caption{Relative ionic population of the most abundant atomic elements contributing to the opacities (in Fig. \ref{fig:opac_compare}) that are driving different layers of the model atmosphere.}
  \label{fig:ion_frac}
\end{figure*}

\begin{figure}
   \centering
   \includegraphics[width=0.7\textwidth, angle=0]{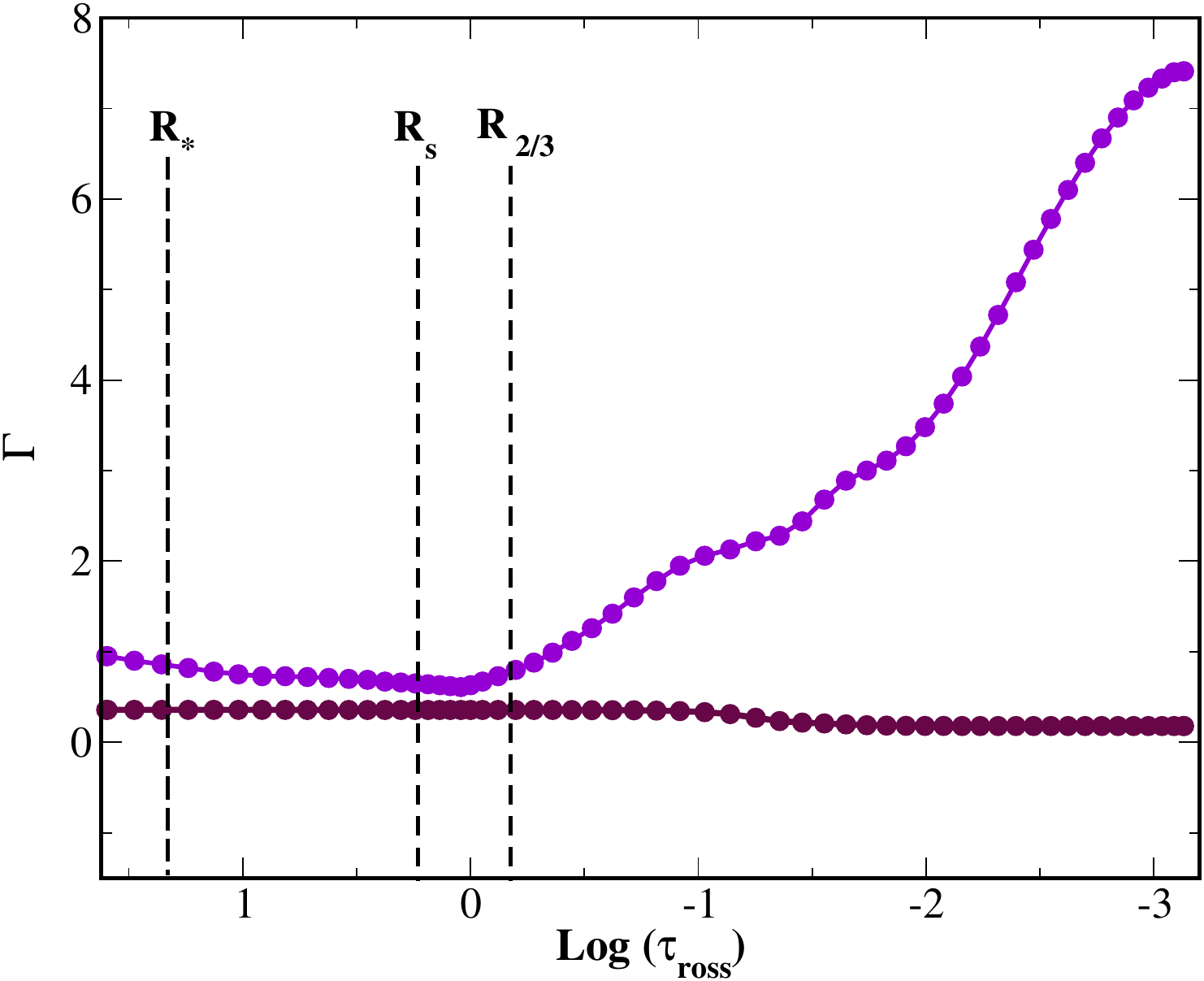}
   \caption{Radiative acceleration ($\Gamma_{rad}$, in \textit{violet}) and electron-scattering (Thomson) acceleration ($\Gamma_{elec}$, in \textit{maroon}) relative to the gravitational acceleration in the stratified layers of the atmospheric model are shown.}
   \label{fig:gamma_compare}
\end{figure}

\begin{figure*}
  \gridline{
    \fig{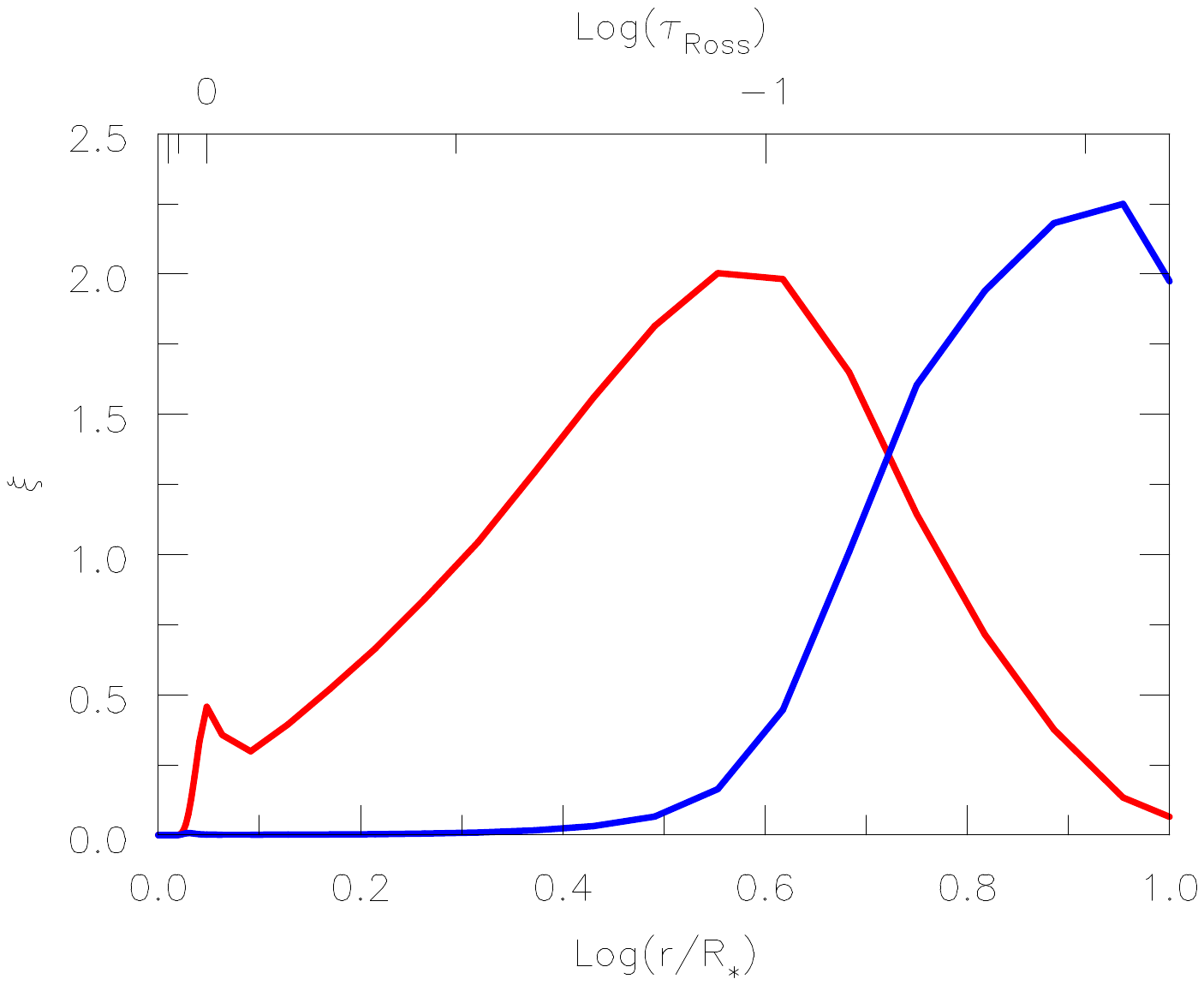}{0.5\linewidth}{(a) Inner region}
    \fig{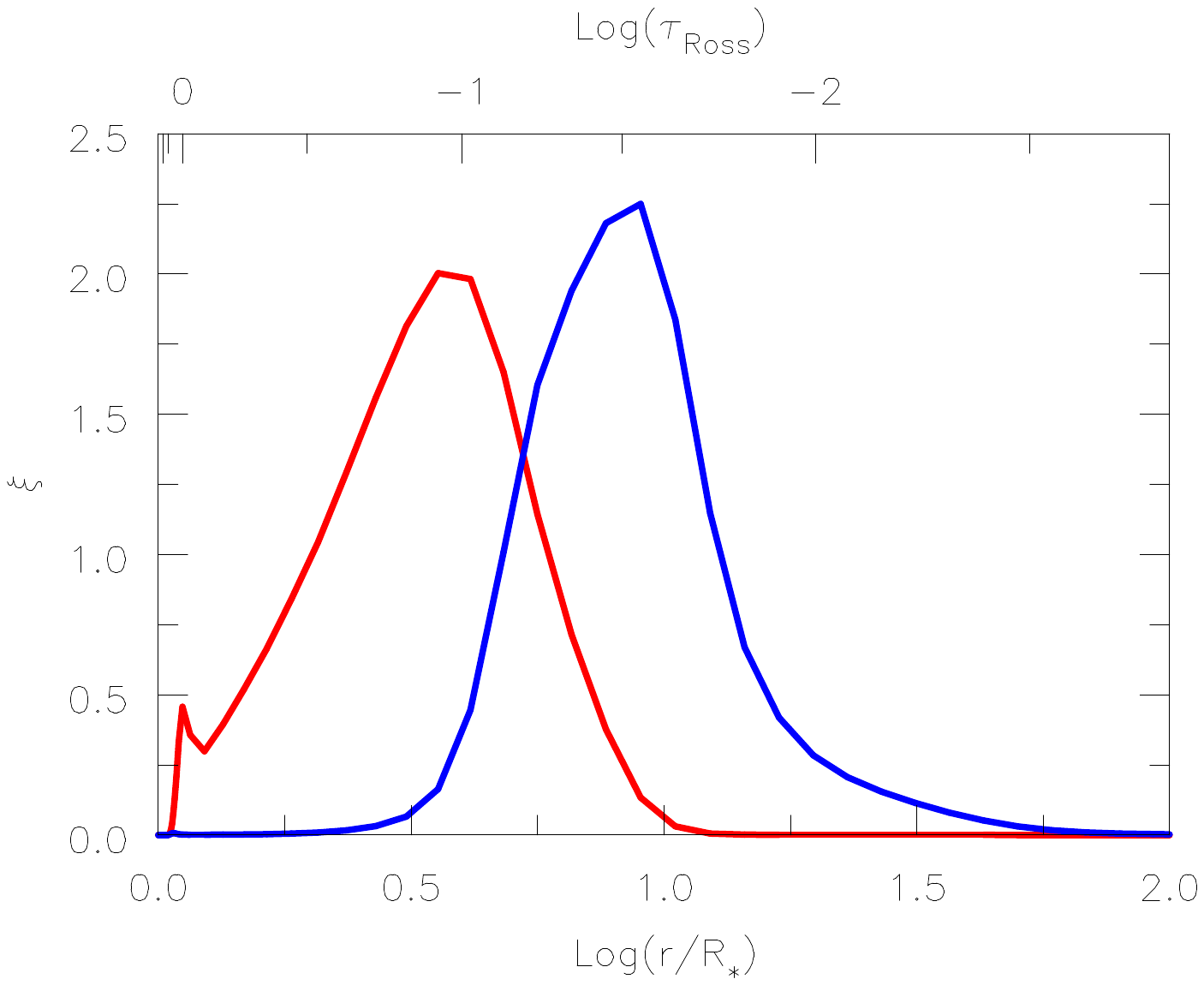}{0.5\linewidth}{(b) Overall region}
  }
  \caption{Emissivity plots showing the line formation region of C\,$\textsc{iv}$ $\lambda\lambda$5802-12 (in \textit{red}), and C\,$\textsc{iii}$ $\lambda$5696 (in \textit{blue}) in the model atmosphere.}
  \label{fig:emissivity_compare}
\end{figure*}

We know that the emission lines affected by the respective line opacities are produced across different regions of the stellar winds. Meanwhile, the continuum opacity (which is independent of the wavelength) is affected by different transitions (due to bound-free, free-free, and electron-scattering processes) that occur across the entire wind. \\
The mean-Rosseland and mean-flux opacities (see Fig. \ref{fig:opac_compare}) show the same nature up to the sonic point. We note (in Fig. \ref{fig:ion_frac} (d)) that the inner sub-sonic winds are driven by the Fe\,$\textsc{vii-ix}$ bound-bound transitions. This is further confirmed from the result (Fig. \ref{fig:gamma_compare}) that the stellar surface ($\tau_{ross}\sim$20) and the sub-photospheric layers lie close to the Eddington limit. The Fe-opacity peak must be located deep inside the convective layer below the stellar surface ($R_{\ast}$). It inflates the outer radiative envelopes and increases the stellar radius \citep{2013A&A...560A...6G}. This shifts the position of the sonic point ($R_{s}$) towards lower optical depths. We find a sudden change in both mean-Rosseland and mean-flux opacities around a certain optical depth ($\tau_{ross}\sim0.1-0.01$) that we claim to be the cold opacity bump (hereafter, COB) where most of the optically thin transitions occur. The electron scattering (Thomson) opacity has no significant influence on the wind-driving mechanism compared to the mean-flux opacity. \\ 
The emissivity of emission lines represents the line formation region. From Fig. \ref{fig:emissivity_compare}, we see that the C\,$\textsc{iv}$ $\lambda\lambda$\,5802-12 originates at a much deeper optical depth than the sonic point ($R_{s}$). A maximum in the emissivity of C\,$\textsc{iv}$ $\lambda\lambda$\,5802-12 is noted around $\tau_{ross}\sim0.1$ and for C\,$\textsc{iii}$ $\lambda$5696 around $\tau_{ross}\sim0.03$. This indicates that the corresponding upper-level UV transitions, i.e. C\,$\textsc{iv}$ $\lambda$193 \& C\,$\textsc{iii}$ $\lambda$312 exhibit strong opacities around the same region ($\tau_{ross}\sim0.1-0.01$).
However, we find that the optically thin wind ($\tau_{ross}\sim0.1-0.01$) is being driven by the He\,$\textsc{ii}$ and C\,$\textsc{iv}$ opacities. This is supported by the increased excitation of $\mathrm{C^{+3}}$ and $\mathrm{He^{+1}}$ ions as seen from the relative population density of the respective ions (see Fig.\ref{fig:ion_frac} (a) and (b)) across the same region. Hence, the mean-flux opacity in those parts of the supersonic winds is mainly contributed by the He\,$\textsc{ii}$ and C\,$\textsc{iv}$ transitions rather than O\,$\textsc{iii}-\textsc{iv}$ and Fe\,$\textsc{iii}-\textsc{vi}$ opacities. It is around the same optical depths we find that $P_{gas}/P_{rad}$ curve shows a sudden shift in the declining slope (see Fig. \ref{fig:Pgas_Prad}(b)).
Therefore, the sudden increase in the radiation pressure is due to the enhanced opacity of the C\,$\textsc{iv}$ and He\,$\textsc{ii}$ UV-transitions which drive the wind in those layers of the atmosphere. \\
\section{Discussion}\label{sect:discussion}
We find that WR\,135 primarily oscillates with the fundamental frequency (see Sec. \ref{subsec:harmonics}). For a rotating star (with $R_{\ast}=6.5\,R_{\odot}$ and $M=19\,M_{\odot}$), the frequency is close to the object's breakup velocity (627\,$kms^{-1}$). Also, for a compact binary companion (of mass 10\,$M_{\odot}$) to exhibit such a period, it needs to be at a separation of 8.7\,$R_{\odot}$ which lies close to the photosphere of the object. Therefore, we safely discard the concept of stellar rotation or the role of a binary companion responsible for the observed variability. Across all the epochs (TESS Sectors: 41, 54, and 55), the primary frequency of the pulsation is complemented by multiple harmonics. From optical photometric lightcurves, we detect harmonics up to $6^{th}$ order, while the higher order such as $8^{th}$ order is detected from the varying emission line strengths of optically thin transitions sampled at equal time intervals. In Sec. \ref{subsec:harmonics}, we find that the line strengths of both C\,$\textsc{iii}$ $\lambda$5696 and C\,$\textsc{iv}$ $\lambda\lambda$\,5802-12 show the same period of variability with a zero phase difference. Such temporal coherence is due to an equal amount of change in the respective line opacities. In Fig. \ref{fig:emissivity_compare} (a), we see that the emission lines are formed in the inner layers of the stellar atmosphere and are more prone to higher-order harmonics. In Sec. \ref{subsec:clump}, we mentioned that the smaller clumps ($f_{VFF}=0.27-0.3$) are participating in the C\,$\textsc{iii}$ $\lambda$5696 and C\,$\textsc{iv}$ $\lambda\lambda$\,5802-12 transitions. These clumps dominate the corresponding line formation region (see Fig. \ref{fig:emissivity_compare}) which lies in the optically thin supersonic winds (see Fig.\ref{fig:vel_compare}). The cold opacity bump (see Fig. \ref{fig:opac_compare}) around the same optical depth further drives the outer supersonic winds where these clumps grow bigger with the increasing radius ($f_{VFF}=0.2$) and affect the overall nature of the optically thin winds. As the clumps of smaller size scatter less flux, the higher$-$order frequencies (with low amplitude) are easily detected from the line strength variation of C\,$\textsc{iv}$ and C\,$\textsc{iii}$ recombination lines. While the bigger clumps participate in the bound-bound transitions (such as the neutral He-transitions: He\,$\textsc{i}$ $\lambda$5876) and the overall flux (line and continuum opacity) from the star. Therefore clumps of different sizes exist at different optical depths of the winds while contributing to the opacities and pulsating under different harmonics of the fundamental frequency (2.73\,$day^{-1}$). However, all types of clumps participate in the stochastic variability of WR\,135 \citep{2022ApJ...925...79L}.

The pulsations in WR stars (e.g., WR\,123) are primarily attributed to two mechanisms: gravity modes from the $\kappa$-mechanism \citep{2006MNRAS.368L..57T} and strange modes resulting from SMIs \citep{2006A&A...453L..35D}. The $\kappa$-mechanism driven pulsations arise due to an opacity peak (at T=$10^{6.25}$\,K due to Fe L-shell transitions) deep within the hydrogen-depleted star \citep{2006MNRAS.368L..57T} and align quite well with the observed periods. However, this mechanism faces challenges due to assumptions about the star's smaller radius compared to the spectroscopically derived values. In contrast, pulsation due to strange modes can occur in stars with suitable radius and hydrogen abundance (e.g., X=0.35 for WR 123) which is higher than typically observed in cWR stars.  Nevertheless, \citet{2008ASPC..391..307G} showed that in a non-linear regime, shock-induced inflations from SMIs can explain the observed pulsation periods in a hydrogen-depleted Helium-Zero Age Main Sequence\,(He-ZAMS), resembling cWR stars. 

Therefore, observed pulsations in WR\,135 could be strange-mode acoustic waves \citep{2008ASPC..391..307G} caused by the SMIs due to the radiative envelope inflation \citep{2013A&A...560A...6G} by the strong Fe-opacity (from Fe\,$\textsc{vii}$-$\textsc{ix}$ transitions) located below the stellar surface \citep{2020MNRAS.491.4406S}. This is supported by the fact that the sub-surface layers (R$<R_{\ast}$) are close to the Eddington limit (see Fig. \ref{fig:gamma_compare}). Also, the density inhomogeneities (i.e. clumps) in the stellar winds could have originated due to the same non-linear SMIs. The envelope inflation noted by \citet{2012A&A...538A..40G} indicates a density inversion in a pure hydrostatic He-ZAMS star. In contrast, our CMFGEN models use a hydrodynamic approach in an optically thin atmosphere, where emission lines and the continuum emerge from the photosphere ($\tau_{ross}$ = 2/3) and are least influenced by the inner hydrostatic layers ($\tau_{ross} < 20$). Thus, our models cannot accurately determine the density stratification between the core ($R_{c}$) and the effective stellar surface ($R_{\ast}$).

On the contrary, we can say that the pulsations could be self-excited waves generated at the base of the winds due to velocity perturbation. \citet{1988ApJ...335..914O} mentioned that the sound waves generated near the subsonic region of the wind get amplified in the supersonic winds by the LDIs. Beyond the sonic point, the strong radiation pressure transforms these waves into inward mode radiative-acoustic waves \citep{1980ApJ...242.1183A}. The outward-mode waves travel faster than the characteristic flow velocity. As these modes reach the optically thin supersonic region ($\tau_{ross}=0.1-0.01$), they are easily detected from the photometric and spectroscopic temporal observations. 

\citet{2016A&A...590A..12G} showed that the strange mode radial pulsations are damped in the case of pure Helium-MS stars with masses greater than 14\,$M_{\odot}$ and the amplitude of pulsations decrease with increasing mass \citep{2008ASPC..391..307G}. WR\,135 with M$\sim$21\,$M_{\odot}$ and C/He=0.25 chemically resemble a He-ZAMS star. Therefore, even if it can host radial strange-mode pulsations the corresponding amplitude would be small, which is neither seen in our results (see Fig.\ref{fig:TESS_pow_spec}) nor detected by \citet{2021MNRAS.502.5038N}. Alternatively, the pulsation can be of non-radial type. Also, we find that the fundamental pulsating period (8.8\,hours) of WR\,135 is quite similar to that detected (9.8\,hours) for WR 123 (WN8-type) from photometric and spectroscopic temporal observations by \citet{2005ApJ...634L.109L} and \citet{2011A&A...530A.151C} respectively. The slight difference between their fundamental frequencies can be attributed to the respective MS progenitors of the objects.
\section{Conclusions}\label{sect:conclusion}
In this work, we investigated the driving source of the stellar winds of WR\,135 which also exhibits pulsations. From the LSP analysis of the photometric and spectroscopic temporal datasets, we identified the higher-order harmonics of the pulsation up to the $8^{th}$ order. To understand the physical conditions in the stellar atmosphere, we used CMFGEN to tailor the best-fitted spectroscopic models to the observed data. We find that the radiation pressure dominates the stellar winds that maintain a hydrodynamic outflow. We showed that the opacity due to Fe\,$\textsc{vii-ix}$ transitions are sufficient enough to launch the inner subsonic winds that are close to the Eddington limit. In contrast, the outer supersonic winds are driven by the He\,$\textsc{ii}$ and C\,$\textsc{iv}$ opacities. The same Fe-opacity could also be responsible for the SMIs producing wind inhomogeneities (clumps) and the SMPs producing different harmonics of the pulsation. We find that the gas producing the unblended optical emission lines (i.e., C\,$\textsc{iv}$ and C\,$\textsc{iii}$) occupy a substantial volume of the stellar winds, suggesting that clumps of the same size are involved in both transitions \citep{2000AJ....120.3201L}. This suggests a strong dependence of the detected harmonics on the size of the clumps in the stellar winds. The smaller clumps participate in the C\,$\textsc{iv}$ and C\,$\textsc{iii}$ recombination transitions oscillating under higher-order harmonics. With increasing radius (in the less accelerated regions of the stellar winds), the clumps grow larger and dominate the outer winds. Such clumps mainly participate in the neutral He\,$\textsc{i}$ transitions, and the overall wind variability.

\section{Acknowledgements}
We thank our reviewer for the constructive suggestions and feedback that enhanced this research work. We acknowledge S. N. Bose National Centre for Basic Sciences under the Department of Science and Technology (DST), Govt. of India, for providing the necessary support to conduct research work. We thank the staff of IAO, Hanle and CREST, Hosakote, who made these observations possible. The facilities at IAO and CREST are operated by the Indian Institute of Astrophysics, Bangalore. The CFHT Science Archive contains data and meta-data provided by the Canada-France-Hawaii Telescope. We acknowledge the use of TESS High Level Science Products (HLSP) produced by the Quick-Look Pipeline (QLP) at the TESS Science Office at MIT, which are publicly available from the Mikulski Archive for Space Telescopes (MAST). Funding for the TESS mission is provided by NASA's Science Mission directorate. This research made use of Lightkurve, a Python package for Kepler and TESS data analysis (Lightkurve Collaboration, 2018). This research has made use of the VizieR catalog access tool, CDS, Strasbourg, France. This publication makes use of data products from the Two Micron All Sky Survey, Wide-field Infrared Survey Explorer and Spitzer (GLIMPSE-survey) which is a joint project of the University of Massachusetts and the Infrared Processing and Analysis Center/California Institute of Technology, funded by the National Aeronautics and Space Administration and the National Science Foundation. This work has made use of data from the European Space Agency (ESA) mission
{\it Gaia} (\url{https://www.cosmos.esa.int/gaia}), processed by the {\it Gaia}
Data Processing and Analysis Consortium (DPAC,
\url{https://www.cosmos.esa.int/web/gaia/dpac/consortium}). Funding for the DPAC
has been provided by national institutions, in particular, the institutions
participating in the {\it Gaia} Multilateral Agreement.

\bibliography{References}{}

\begin{thebibliography}{}
\expandafter\ifx\csname natexlab\endcsname\relax\def\natexlab#1{#1}\fi
\providecommand{\url}[1]{\href{#1}{#1}}
\providecommand{\dodoi}[1]{doi:~\href{http://doi.org/#1}{\nolinkurl{#1}}}
\providecommand{\doeprint}[1]{\href{http://ascl.net/#1}{\nolinkurl{http://ascl.net/#1}}}
\providecommand{\doarXiv}[1]{\href{https://arxiv.org/abs/#1}{\nolinkurl{https://arxiv.org/abs/#1}}}

\bibitem[{{Aadland} {et~al.}(2022){Aadland}, {Massey}, {Hillier}, \& {Morrell}}]{2022ApJ...924...44A}
{Aadland}, E., {Massey}, P., {Hillier}, D.~J., \& {Morrell}, N. 2022, \apj, 924, 44, \dodoi{10.3847/1538-4357/ac3426}

\bibitem[{{Abbott}(1980)}]{1980ApJ...242.1183A}
{Abbott}, D.~C. 1980, \apj, 242, 1183, \dodoi{10.1086/158550}

\bibitem[{{Ardila} {et~al.}(2010{\natexlab{a}}){Ardila}, {Van Dyk}, {Makowiecki}, {Stauffer}, {Song}, {Rho}, {Fajardo-Acosta}, {Hoard}, \& {Wachter}}]{2010ApJS..191..301A}
{Ardila}, D.~R., {Van Dyk}, S.~D., {Makowiecki}, W., {et~al.} 2010{\natexlab{a}}, \apjs, 191, 301, \dodoi{10.1088/0067-0049/191/2/301}

\bibitem[{{Ardila} {et~al.}(2010{\natexlab{b}}){Ardila}, {van Dyk}, {Makowiecki}, {Stauffer}, {Song}, {Rho}, {Fajardo-Acosta}, {Hoard}, \& {Wachter}}]{2010yCat..21910301A}
{Ardila}, D.~R., {van Dyk}, S.~D., {Makowiecki}, W., {et~al.} 2010{\natexlab{b}}, {VizieR Online Data Catalog: Spitzer Atlas of Stellar Spectra (SASS) (Ardila+, 2010)}, VizieR On-line Data Catalog: J/ApJS/191/301. Originally published in: 2010ApJS..191..301A, \dodoi{10.26093/cds/vizier.21910301}

\bibitem[{{Astropy Collaboration} {et~al.}(2013){Astropy Collaboration}, {Robitaille}, {Tollerud}, {Greenfield}, {Droettboom}, {Bray}, {Aldcroft}, {Davis}, {Ginsburg}, {Price-Whelan}, {Kerzendorf}, {Conley}, {Crighton}, {Barbary}, {Muna}, {Ferguson}, {Grollier}, {Parikh}, {Nair}, {Unther}, {Deil}, {Woillez}, {Conseil}, {Kramer}, {Turner}, {Singer}, {Fox}, {Weaver}, {Zabalza}, {Edwards}, {Azalee Bostroem}, {Burke}, {Casey}, {Crawford}, {Dencheva}, {Ely}, {Jenness}, {Labrie}, {Lim}, {Pierfederici}, {Pontzen}, {Ptak}, {Refsdal}, {Servillat}, \& {Streicher}}]{2013A&A...558A..33A}
{Astropy Collaboration}, {Robitaille}, T.~P., {Tollerud}, E.~J., {et~al.} 2013, \aap, 558, A33, \dodoi{10.1051/0004-6361/201322068}

\bibitem[{{Astropy Collaboration} {et~al.}(2018){Astropy Collaboration}, {Price-Whelan}, {Sip{\H{o}}cz}, {G{\"u}nther}, {Lim}, {Crawford}, {Conseil}, {Shupe}, {Craig}, {Dencheva}, {Ginsburg}, {VanderPlas}, {Bradley}, {P{\'e}rez-Su{\'a}rez}, {de Val-Borro}, {Aldcroft}, {Cruz}, {Robitaille}, {Tollerud}, {Ardelean}, {Babej}, {Bach}, {Bachetti}, {Bakanov}, {Bamford}, {Barentsen}, {Barmby}, {Baumbach}, {Berry}, {Biscani}, {Boquien}, {Bostroem}, {Bouma}, {Brammer}, {Bray}, {Breytenbach}, {Buddelmeijer}, {Burke}, {Calderone}, {Cano Rodr{\'\i}guez}, {Cara}, {Cardoso}, {Cheedella}, {Copin}, {Corrales}, {Crichton}, {D'Avella}, {Deil}, {Depagne}, {Dietrich}, {Donath}, {Droettboom}, {Earl}, {Erben}, {Fabbro}, {Ferreira}, {Finethy}, {Fox}, {Garrison}, {Gibbons}, {Goldstein}, {Gommers}, {Greco}, {Greenfield}, {Groener}, {Grollier}, {Hagen}, {Hirst}, {Homeier}, {Horton}, {Hosseinzadeh}, {Hu}, {Hunkeler}, {Ivezi{\'c}}, {Jain}, {Jenness}, {Kanarek}, {Kendrew}, {Kern}, {Kerzendorf}, {Khvalko}, {King}, {Kirkby}, {Kulkarni},
  {Kumar}, {Lee}, {Lenz}, {Littlefair}, {Ma}, {Macleod}, {Mastropietro}, {McCully}, {Montagnac}, {Morris}, {Mueller}, {Mumford}, {Muna}, {Murphy}, {Nelson}, {Nguyen}, {Ninan}, {N{\"o}the}, {Ogaz}, {Oh}, {Parejko}, {Parley}, {Pascual}, {Patil}, {Patil}, {Plunkett}, {Prochaska}, {Rastogi}, {Reddy Janga}, {Sabater}, {Sakurikar}, {Seifert}, {Sherbert}, {Sherwood-Taylor}, {Shih}, {Sick}, {Silbiger}, {Singanamalla}, {Singer}, {Sladen}, {Sooley}, {Sornarajah}, {Streicher}, {Teuben}, {Thomas}, {Tremblay}, {Turner}, {Terr{\'o}n}, {van Kerkwijk}, {de la Vega}, {Watkins}, {Weaver}, {Whitmore}, {Woillez}, {Zabalza}, \& {Astropy Contributors}}]{2018AJ....156..123A}
{Astropy Collaboration}, {Price-Whelan}, A.~M., {Sip{\H{o}}cz}, B.~M., {et~al.} 2018, \aj, 156, 123, \dodoi{10.3847/1538-3881/aabc4f}

\bibitem[{{Astropy Collaboration} {et~al.}(2022){Astropy Collaboration}, {Price-Whelan}, {Lim}, {Earl}, {Starkman}, {Bradley}, {Shupe}, {Patil}, {Corrales}, {Brasseur}, {N{\"o}the}, {Donath}, {Tollerud}, {Morris}, {Ginsburg}, {Vaher}, {Weaver}, {Tocknell}, {Jamieson}, {van Kerkwijk}, {Robitaille}, {Merry}, {Bachetti}, {G{\"u}nther}, {Aldcroft}, {Alvarado-Montes}, {Archibald}, {B{\'o}di}, {Bapat}, {Barentsen}, {Baz{\'a}n}, {Biswas}, {Boquien}, {Burke}, {Cara}, {Cara}, {Conroy}, {Conseil}, {Craig}, {Cross}, {Cruz}, {D'Eugenio}, {Dencheva}, {Devillepoix}, {Dietrich}, {Eigenbrot}, {Erben}, {Ferreira}, {Foreman-Mackey}, {Fox}, {Freij}, {Garg}, {Geda}, {Glattly}, {Gondhalekar}, {Gordon}, {Grant}, {Greenfield}, {Groener}, {Guest}, {Gurovich}, {Handberg}, {Hart}, {Hatfield-Dodds}, {Homeier}, {Hosseinzadeh}, {Jenness}, {Jones}, {Joseph}, {Kalmbach}, {Karamehmetoglu}, {Ka{\l}uszy{\'n}ski}, {Kelley}, {Kern}, {Kerzendorf}, {Koch}, {Kulumani}, {Lee}, {Ly}, {Ma}, {MacBride}, {Maljaars}, {Muna}, {Murphy}, {Norman},
  {O'Steen}, {Oman}, {Pacifici}, {Pascual}, {Pascual-Granado}, {Patil}, {Perren}, {Pickering}, {Rastogi}, {Roulston}, {Ryan}, {Rykoff}, {Sabater}, {Sakurikar}, {Salgado}, {Sanghi}, {Saunders}, {Savchenko}, {Schwardt}, {Seifert-Eckert}, {Shih}, {Jain}, {Shukla}, {Sick}, {Simpson}, {Singanamalla}, {Singer}, {Singhal}, {Sinha}, {Sip{\H{o}}cz}, {Spitler}, {Stansby}, {Streicher}, {{\v{S}}umak}, {Swinbank}, {Taranu}, {Tewary}, {Tremblay}, {de Val-Borro}, {Van Kooten}, {Vasovi{\'c}}, {Verma}, {de Miranda Cardoso}, {Williams}, {Wilson}, {Winkel}, {Wood-Vasey}, {Xue}, {Yoachim}, {Zhang}, {Zonca}, \& {Astropy Project Contributors}}]{2022ApJ...935..167A}
{Astropy Collaboration}, {Price-Whelan}, A.~M., {Lim}, P.~L., {et~al.} 2022, \apj, 935, 167, \dodoi{10.3847/1538-4357/ac7c74}

\bibitem[{{Bailer-Jones} {et~al.}(2018){Bailer-Jones}, {Rybizki}, {Fouesneau}, {Mantelet}, \& {Andrae}}]{2018AJ....156...58B}
{Bailer-Jones}, C.~A.~L., {Rybizki}, J., {Fouesneau}, M., {Mantelet}, G., \& {Andrae}, R. 2018, \aj, 156, 58, \dodoi{10.3847/1538-3881/aacb21}

\bibitem[{{Cantiello} {et~al.}(2009){Cantiello}, {Langer}, {Brott}, {de Koter}, {Shore}, {Vink}, {Voegler}, {Lennon}, \& {Yoon}}]{2009A&A...499..279C}
{Cantiello}, M., {Langer}, N., {Brott}, I., {et~al.} 2009, \aap, 499, 279, \dodoi{10.1051/0004-6361/200911643}

\bibitem[{{Castor} {et~al.}(1975){Castor}, {Abbott}, \& {Klein}}]{1975ApJ...195..157C}
{Castor}, J.~I., {Abbott}, D.~C., \& {Klein}, R.~I. 1975, \apj, 195, 157, \dodoi{10.1086/153315}

\bibitem[{{Chen{\'e}} \& {St-Louis}(2011)}]{2011ApJ...736..140C}
{Chen{\'e}}, A.~N., \& {St-Louis}, N. 2011, \apj, 736, 140, \dodoi{10.1088/0004-637X/736/2/140}

\bibitem[{{Chen{\'e}} {et~al.}(2020){Chen{\'e}}, {St-Louis}, {Moffat}, \& {Gayley}}]{2020ApJ...903..113C}
{Chen{\'e}}, A.-N., {St-Louis}, N., {Moffat}, A. F.~J., \& {Gayley}, K.~G. 2020, \apj, 903, 113, \dodoi{10.3847/1538-4357/abba24}

\bibitem[{{Chen{\'e}} {et~al.}(2011){Chen{\'e}}, {Foellmi}, {Marchenko}, {St-Louis}, {Moffat}, {Ballereau}, {Chauville}, {Zorec}, \& {Poteet}}]{2011A&A...530A.151C}
{Chen{\'e}}, A.~N., {Foellmi}, C., {Marchenko}, S.~V., {et~al.} 2011, \aap, 530, A151, \dodoi{10.1051/0004-6361/201116567}

\bibitem[{{Cutri} {et~al.}(2003){Cutri}, {Skrutskie}, {Van Dyk}, {Beichman}, {Carpenter}, {Chester}, {Cambresy}, {Evans}, {Fowler}, {Gizis}, {Howard}, {Huchra}, {Jarrett}, {Kopan}, {Kirkpatrick}, {Light}, {Marsh}, {McCallon}, {Schneider}, {Stiening}, {Sykes}, {Weinberg}, {Wheaton}, {Wheelock}, \& {Zacarias}}]{vizierII246}
{Cutri}, R., {Skrutskie}, M., {Van Dyk}, S., {et~al.} 2003, {2MASS All-Sky Catalog of Point Sources}, Version 17-Apr-2020 (last modified),  Centre de Donnees astronomique de Strasbourg (CDS)

\bibitem[{{Cutri} \& {et al.}(2012)}]{2012yCat.2311....0C}
{Cutri}, R.~M., \& {et al.} 2012, {VizieR Online Data Catalog: WISE All-Sky Data Release (Cutri+ 2012)}, VizieR On-line Data Catalog: II/311. Originally published in: 2012wise.rept....1C

\bibitem[{{Cutri} {et~al.}(2012){Cutri}, {Wright}, {Conrow}, {Bauer}, {Benford}, {Brandenburg}, {Dailey}, {Eisenhardt}, {Evans}, {Fajardo-Acosta}, {Fowler}, {Gelino}, {Grillmair}, {Harbut}, {Hoffman}, {Jarrett}, {Kirkpatrick}, {Leisawitz}, {Liu}, {Mainzer}, {Marsh}, {Masci}, {McCallon}, {Padgett}, {Ressler}, {Royer}, {Skrutskie}, {Stanford}, {Wyatt}, {Tholen}, {Tsai}, {Wachter}, {Wheelock}, {Yan}, {Alles}, {Beck}, {Grav}, {Masiero}, {McCollum}, {McGehee}, {Papin}, \& {Wittman}}]{2012wise.rept....1C}
{Cutri}, R.~M., {Wright}, E.~L., {Conrow}, T., {et~al.} 2012, {Explanatory Supplement to the WISE All-Sky Data Release Products}, Explanatory Supplement to the WISE All-Sky Data Release Products

\bibitem[{{de la Chevroti{\`e}re} {et~al.}(2014){de la Chevroti{\`e}re}, {St-Louis}, {Moffat}, \& {MiMeS Collaboration}}]{2014ApJ...781...73D}
{de la Chevroti{\`e}re}, A., {St-Louis}, N., {Moffat}, A.~F.~J., \& {MiMeS Collaboration}. 2014, \apj, 781, 73, \dodoi{10.1088/0004-637X/781/2/73}

\bibitem[{{Decleir} {et~al.}(2022){Decleir}, {Gordon}, {Andrews}, {Clayton}, {Cushing}, {Misselt}, {Pendleton}, {Rayner}, {Vacca}, \& {Whittet}}]{2022ApJ...930...15D}
{Decleir}, M., {Gordon}, K.~D., {Andrews}, J.~E., {et~al.} 2022, \apj, 930, 15, \dodoi{10.3847/1538-4357/ac5dbe}

\bibitem[{{Donati} {et~al.}(1997){Donati}, {Semel}, {Carter}, {Rees}, \& {Collier Cameron}}]{1997MNRAS.291..658D}
{Donati}, J.~F., {Semel}, M., {Carter}, B.~D., {Rees}, D.~E., \& {Collier Cameron}, A. 1997, \mnras, 291, 658, \dodoi{10.1093/mnras/291.4.658}

\bibitem[{{Dorfi} {et~al.}(2006){Dorfi}, {Gautschy}, \& {Saio}}]{2006A&A...453L..35D}
{Dorfi}, E.~A., {Gautschy}, A., \& {Saio}, H. 2006, \aap, 453, L35, \dodoi{10.1051/0004-6361:200600027}

\bibitem[{{Esa}(1997)}]{1997yCat.1239....0E}
{Esa}, . 1997, {VizieR Online Data Catalog: The Hipparcos and Tycho Catalogues (ESA 1997)}, VizieR On-line Data Catalog: I/239. Originally published in: 1997HIP...C......0E

\bibitem[{{Fitzpatrick} {et~al.}(2019){Fitzpatrick}, {Massa}, {Gordon}, {Bohlin}, \& {Clayton}}]{2019ApJ...886..108F}
{Fitzpatrick}, E.~L., {Massa}, D., {Gordon}, K.~D., {Bohlin}, R., \& {Clayton}, G.~C. 2019, \apj, 886, 108, \dodoi{10.3847/1538-4357/ab4c3a}

\bibitem[{{Gaia Collaboration}(2018)}]{2018yCat.1345....0G}
{Gaia Collaboration}. 2018, {VizieR Online Data Catalog: Gaia DR2 (Gaia Collaboration, 2018)}, VizieR On-line Data Catalog: I/345. Originally published in: 2018A\&A...616A...1G; doi:10.5270/esa-ycs, \dodoi{10.26093/cds/vizier.1345}

\bibitem[{{Gaia Collaboration} {et~al.}(2018){Gaia Collaboration}, {Brown}, {Vallenari}, {Prusti}, {de Bruijne}, {Babusiaux}, {Bailer-Jones}, {Biermann}, {Evans}, {Eyer}, {Jansen}, {Jordi}, {Klioner}, {Lammers}, {Lindegren}, {Luri}, {Mignard}, {Panem}, {Pourbaix}, {Randich}, {Sartoretti}, {Siddiqui}, {Soubiran}, {van Leeuwen}, {Walton}, {Arenou}, {Bastian}, {Cropper}, {Drimmel}, {Katz}, {Lattanzi}, {Bakker}, {Cacciari}, {Casta{\~n}eda}, {Chaoul}, {Cheek}, {De Angeli}, {Fabricius}, {Guerra}, {Holl}, {Masana}, {Messineo}, {Mowlavi}, {Nienartowicz}, {Panuzzo}, {Portell}, {Riello}, {Seabroke}, {Tanga}, {Th{\'e}venin}, {Gracia-Abril}, {Comoretto}, {Garcia-Reinaldos}, {Teyssier}, {Altmann}, {Andrae}, {Audard}, {Bellas-Velidis}, {Benson}, {Berthier}, {Blomme}, {Burgess}, {Busso}, {Carry}, {Cellino}, {Clementini}, {Clotet}, {Creevey}, {Davidson}, {De Ridder}, {Delchambre}, {Dell'Oro}, {Ducourant}, {Fern{\'a}ndez-Hern{\'a}ndez}, {Fouesneau}, {Fr{\'e}mat}, {Galluccio}, {Garc{\'\i}a-Torres},
  {Gonz{\'a}lez-N{\'u}{\~n}ez}, {Gonz{\'a}lez-Vidal}, {Gosset}, {Guy}, {Halbwachs}, {Hambly}, {Harrison}, {Hern{\'a}ndez}, {Hestroffer}, {Hodgkin}, {Hutton}, {Jasniewicz}, {Jean-Antoine-Piccolo}, {Jordan}, {Korn}, {Krone-Martins}, {Lanzafame}, {Lebzelter}, {L{\"o}ffler}, {Manteiga}, {Marrese}, {Mart{\'\i}n-Fleitas}, {Moitinho}, {Mora}, {Muinonen}, {Osinde}, {Pancino}, {Pauwels}, {Petit}, {Recio-Blanco}, {Richards}, {Rimoldini}, {Robin}, {Sarro}, {Siopis}, {Smith}, {Sozzetti}, {S{\"u}veges}, {Torra}, {van Reeven}, {Abbas}, {Abreu Aramburu}, {Accart}, {Aerts}, {Altavilla}, {{\'A}lvarez}, {Alvarez}, {Alves}, {Anderson}, {Andrei}, {Anglada Varela}, {Antiche}, {Antoja}, {Arcay}, {Astraatmadja}, {Bach}, {Baker}, {Balaguer-N{\'u}{\~n}ez}, {Balm}, {Barache}, {Barata}, {Barbato}, {Barblan}, {Barklem}, {Barrado}, {Barros}, {Barstow}, {Bartholom{\'e} Mu{\~n}oz}, {Bassilana}, {Becciani}, {Bellazzini}, {Berihuete}, {Bertone}, {Bianchi}, {Bienaym{\'e}}, {Blanco-Cuaresma}, {Boch}, {Boeche}, {Bombrun}, {Borrachero},
  {Bossini}, {Bouquillon}, {Bourda}, {Bragaglia}, {Bramante}, {Breddels}, {Bressan}, {Brouillet}, {Br{\"u}semeister}, {Brugaletta}, {Bucciarelli}, {Burlacu}, {Busonero}, {Butkevich}, {Buzzi}, {Caffau}, {Cancelliere}, {Cannizzaro}, {Cantat-Gaudin}, {Carballo}, {Carlucci}, {Carrasco}, {Casamiquela}, {Castellani}, {Castro-Ginard}, {Charlot}, {Chemin}, {Chiavassa}, {Cocozza}, {Costigan}, {Cowell}, {Crifo}, {Crosta}, {Crowley}, {Cuypers}, {Dafonte}, {Damerdji}, {Dapergolas}, {David}, {David}, {de Laverny}, {De Luise}, {De March}, {de Martino}, {de Souza}, {de Torres}, {Debosscher}, {del Pozo}, {Delbo}, {Delgado}, {Delgado}, {Di Matteo}, {Diakite}, {Diener}, {Distefano}, {Dolding}, {Drazinos}, {Dur{\'a}n}, {Edvardsson}, {Enke}, {Eriksson}, {Esquej}, {Eynard Bontemps}, {Fabre}, {Fabrizio}, {Faigler}, {Falc{\~a}o}, {Farr{\`a}s Casas}, {Federici}, {Fedorets}, {Fernique}, {Figueras}, {Filippi}, {Findeisen}, {Fonti}, {Fraile}, {Fraser}, {Fr{\'e}zouls}, {Gai}, {Galleti}, {Garabato}, {Garc{\'\i}a-Sedano}, {Garofalo},
  {Garralda}, {Gavel}, {Gavras}, {Gerssen}, {Geyer}, {Giacobbe}, {Gilmore}, {Girona}, {Giuffrida}, {Glass}, {Gomes}, {Granvik}, {Gueguen}, {Guerrier}, {Guiraud}, {Guti{\'e}rrez-S{\'a}nchez}, {Haigron}, {Hatzidimitriou}, {Hauser}, {Haywood}, {Heiter}, {Helmi}, {Heu}, {Hilger}, {Hobbs}, {Hofmann}, {Holland}, {Huckle}, {Hypki}, {Icardi}, {Jan{\ss}en}, {Jevardat de Fombelle}, {Jonker}, {Juh{\'a}sz}, {Julbe}, {Karampelas}, {Kewley}, {Klar}, {Kochoska}, {Kohley}, {Kolenberg}, {Kontizas}, {Kontizas}, {Koposov}, {Kordopatis}, {Kostrzewa-Rutkowska}, {Koubsky}, {Lambert}, {Lanza}, {Lasne}, {Lavigne}, {Le Fustec}, {Le Poncin-Lafitte}, {Lebreton}, {Leccia}, {Leclerc}, {Lecoeur-Taibi}, {Lenhardt}, {Leroux}, {Liao}, {Licata}, {Lindstr{\o}m}, {Lister}, {Livanou}, {Lobel}, {L{\'o}pez}, {Managau}, {Mann}, {Mantelet}, {Marchal}, {Marchant}, {Marconi}, {Marinoni}, {Marschalk{\'o}}, {Marshall}, {Martino}, {Marton}, {Mary}, {Massari}, {Matijevi{\v{c}}}, {Mazeh}, {McMillan}, {Messina}, {Michalik}, {Millar}, {Molina}, {Molinaro},
  {Moln{\'a}r}, {Montegriffo}, {Mor}, {Morbidelli}, {Morel}, {Morris}, {Mulone}, {Muraveva}, {Musella}, {Nelemans}, {Nicastro}, {Noval}, {O'Mullane}, {Ord{\'e}novic}, {Ord{\'o}{\~n}ez-Blanco}, {Osborne}, {Pagani}, {Pagano}, {Pailler}, {Palacin}, {Palaversa}, {Panahi}, {Pawlak}, {Piersimoni}, {Pineau}, {Plachy}, {Plum}, {Poggio}, {Poujoulet}, {Pr{\v{s}}a}, {Pulone}, {Racero}, {Ragaini}, {Rambaux}, {Ramos-Lerate}, {Regibo}, {Reyl{\'e}}, {Riclet}, {Ripepi}, {Riva}, {Rivard}, {Rixon}, {Roegiers}, {Roelens}, {Romero-G{\'o}mez}, {Rowell}, {Royer}, {Ruiz-Dern}, {Sadowski}, {Sagrist{\`a} Sell{\'e}s}, {Sahlmann}, {Salgado}, {Salguero}, {Sanna}, {Santana-Ros}, {Sarasso}, {Savietto}, {Schultheis}, {Sciacca}, {Segol}, {Segovia}, {S{\'e}gransan}, {Shih}, {Siltala}, {Silva}, {Smart}, {Smith}, {Solano}, {Solitro}, {Sordo}, {Soria Nieto}, {Souchay}, {Spagna}, {Spoto}, {Stampa}, {Steele}, {Steidelm{\"u}ller}, {Stephenson}, {Stoev}, {Suess}, {Surdej}, {Szabados}, {Szegedi-Elek}, {Tapiador}, {Taris}, {Tauran}, {Taylor},
  {Teixeira}, {Terrett}, {Teyssandier}, {Thuillot}, {Titarenko}, {Torra Clotet}, {Turon}, {Ulla}, {Utrilla}, {Uzzi}, {Vaillant}, {Valentini}, {Valette}, {van Elteren}, {Van Hemelryck}, {van Leeuwen}, {Vaschetto}, {Vecchiato}, {Veljanoski}, {Viala}, {Vicente}, {Vogt}, {von Essen}, {Voss}, {Votruba}, {Voutsinas}, {Walmsley}, {Weiler}, {Wertz}, {Wevers}, {Wyrzykowski}, {Yoldas}, {{\v{Z}}erjal}, {Ziaeepour}, {Zorec}, {Zschocke}, {Zucker}, {Zurbach}, \& {Zwitter}}]{2018A&A...616A...1G}
{Gaia Collaboration}, {Brown}, A.~G.~A., {Vallenari}, A., {et~al.} 2018, \aap, 616, A1, \dodoi{10.1051/0004-6361/201833051}

\bibitem[{{Glatzel}(1994)}]{1994MNRAS.271...66G}
{Glatzel}, W. 1994, \mnras, 271, 66, \dodoi{10.1093/mnras/271.1.66}

\bibitem[{{Glatzel}(2008)}]{2008ASPC..391..307G}
{Glatzel}, W. 2008, in Astronomical Society of the Pacific Conference Series, Vol. 391, Hydrogen-Deficient Stars, ed. A.~{Werner} \& T.~{Rauch}, 307

\bibitem[{{Gordon} {et~al.}(2009){Gordon}, {Cartledge}, \& {Clayton}}]{2009ApJ...705.1320G}
{Gordon}, K.~D., {Cartledge}, S., \& {Clayton}, G.~C. 2009, \apj, 705, 1320, \dodoi{10.1088/0004-637X/705/2/1320}

\bibitem[{{Gordon} {et~al.}(2023){Gordon}, {Clayton}, {Decleir}, {Fitzpatrick}, {Massa}, {Misselt}, \& {Tollerud}}]{2023ApJ...950...86G}
{Gordon}, K.~D., {Clayton}, G.~C., {Decleir}, M., {et~al.} 2023, \apj, 950, 86, \dodoi{10.3847/1538-4357/accb59}

\bibitem[{{Gordon} {et~al.}(2021){Gordon}, {Misselt}, {Bouwman}, {Clayton}, {Decleir}, {Hines}, {Pendleton}, {Rieke}, {Smith}, \& {Whittet}}]{2021ApJ...916...33G}
{Gordon}, K.~D., {Misselt}, K.~A., {Bouwman}, J., {et~al.} 2021, \apj, 916, 33, \dodoi{10.3847/1538-4357/ac00b7}

\bibitem[{{Gr{\"a}fener} {et~al.}(2012){Gr{\"a}fener}, {Owocki}, \& {Vink}}]{2012A&A...538A..40G}
{Gr{\"a}fener}, G., {Owocki}, S.~P., \& {Vink}, J.~S. 2012, \aap, 538, A40, \dodoi{10.1051/0004-6361/201117497}

\bibitem[{{Gr{\"a}fener} \& {Vink}(2013)}]{2013A&A...560A...6G}
{Gr{\"a}fener}, G., \& {Vink}, J.~S. 2013, \aap, 560, A6, \dodoi{10.1051/0004-6361/201321914}

\bibitem[{{Gr{\"a}fener} {et~al.}(2011){Gr{\"a}fener}, {Vink}, {de Koter}, \& {Langer}}]{2011A&A...535A..56G}
{Gr{\"a}fener}, G., {Vink}, J.~S., {de Koter}, A., \& {Langer}, N. 2011, \aap, 535, A56, \dodoi{10.1051/0004-6361/201116701}

\bibitem[{{Grassitelli} {et~al.}(2016){Grassitelli}, {Chen{\'e}}, {Sanyal}, {Langer}, {St-Louis}, {Bestenlehner}, \& {Fossati}}]{2016A&A...590A..12G}
{Grassitelli}, L., {Chen{\'e}}, A.~N., {Sanyal}, D., {et~al.} 2016, \aap, 590, A12, \dodoi{10.1051/0004-6361/201527873}

\bibitem[{{Hillier}(1989)}]{1989ApJ...347..392H}
{Hillier}, D.~J. 1989, \apj, 347, 392

\bibitem[{{Hillier}(2003)}]{2003IAUS..212...70H}
{Hillier}, D.~J. 2003, in A Massive Star Odyssey: From Main Sequence to Supernova, ed. K.~{van der Hucht}, A.~{Herrero}, \& C.~{Esteban}, Vol. 212, 70

\bibitem[{{Hillier}(2011)}]{2011Ap&SS.336...87H}
---. 2011, \apss, 336, 87, \dodoi{10.1007/s10509-010-0590-9}

\bibitem[{{Hillier}(2012)}]{2012IAUS..282..229H}
{Hillier}, D.~J. 2012, in From Interacting Binaries to Exoplanets: Essential Modeling Tools, ed. M.~T. {Richards} \& I.~{Hubeny}, Vol. 282, 229--234

\bibitem[{{Hillier} \& {Miller}(1998)}]{1998ApJ...496..407H}
{Hillier}, D.~J., \& {Miller}, D.~L. 1998, \apj, 496, 407

\bibitem[{{Huang} {et~al.}(2020){Huang}, {Vanderburg}, {P{\'a}l}, {Sha}, {Yu}, {Fong}, {Fausnaugh}, {Shporer}, {Guerrero}, {Vanderspek}, \& {Ricker}}]{2020RNAAS...4..204H}
{Huang}, C.~X., {Vanderburg}, A., {P{\'a}l}, A., {et~al.} 2020, Research Notes of the American Astronomical Society, 4, 204, \dodoi{10.3847/2515-5172/abca2e}

\bibitem[{Kar {et~al.}(2024)Kar, Das, \& Baug}]{Kar_2024}
Kar, S., Das, R., \& Baug, T. 2024, The Astrophysical Journal, 968, 60, \dodoi{10.3847/1538-4357/ad4131}

\bibitem[{{Kunimoto} {et~al.}(2022){Kunimoto}, {Tey}, {Fong}, {Hesse}, {Shporer}, {Fausnaugh}, {Vanderspek}, \& {Ricker}}]{2022RNAAS...6..236K}
{Kunimoto}, M., {Tey}, E., {Fong}, W., {et~al.} 2022, Research Notes of the American Astronomical Society, 6, 236, \dodoi{10.3847/2515-5172/aca158}

\bibitem[{Kunimoto {et~al.}(2021)Kunimoto, Huang, Tey, Fong, Hesse, Shporer, Guerrero, Fausnaugh, Vanderspek, \& Ricker}]{Kunimoto_2021}
Kunimoto, M., Huang, C., Tey, E., {et~al.} 2021, Research Notes of the AAS, 5, 234, \dodoi{10.3847/2515-5172/ac2ef0}

\bibitem[{{Lasker} {et~al.}(2007){Lasker}, {Lattanzi}, {McLean}, \& {et al.}}]{2007yCat.1305....0L}
{Lasker}, B., {Lattanzi}, M.~G., {McLean}, B.~J., \& {et al.} 2007, {VizieR Online Data Catalog: The Guide Star Catalog, Version 2.3.2 (GSC2.3) (STScI, 2006)}, VizieR On-line Data Catalog: I/305. Originally published in: 2008AJ....136..735L

\bibitem[{{Lasker} {et~al.}(2008){Lasker}, {Lattanzi}, {McLean}, {Bucciarelli}, {Drimmel}, {Garcia}, {Greene}, {Guglielmetti}, {Hanley}, {Hawkins}, {Laidler}, {Loomis}, {Meakes}, {Mignani}, {Morbidelli}, {Morrison}, {Pannunzio}, {Rosenberg}, {Sarasso}, {Smart}, {Spagna}, {Sturch}, {Volpicelli}, {White}, {Wolfe}, \& {Zacchei}}]{2008AJ....136..735L}
{Lasker}, B.~M., {Lattanzi}, M.~G., {McLean}, B.~J., {et~al.} 2008, \aj, 136, 735, \dodoi{10.1088/0004-6256/136/2/735}

\bibitem[{{Lef{\`e}vre} {et~al.}(2005){Lef{\`e}vre}, {Marchenko}, {Moffat}, {Chen{\'e}}, {Smith}, {St-Louis}, {Matthews}, {Kuschnig}, {Guenther}, {Poteet}, {Rucinski}, {Sasselov}, {Walker}, \& {Weiss}}]{2005ApJ...634L.109L}
{Lef{\`e}vre}, L., {Marchenko}, S.~V., {Moffat}, A.~F.~J., {et~al.} 2005, \apjl, 634, L109, \dodoi{10.1086/498393}

\bibitem[{{Lenoir-Craig} {et~al.}(2022){Lenoir-Craig}, {St-Louis}, {Moffat}, {Pablo}, {Handler}, {Kuschnig}, {Popowicz}, {Wade}, \& {Zwintz}}]{2022ApJ...925...79L}
{Lenoir-Craig}, G., {St-Louis}, N., {Moffat}, A. F.~J., {et~al.} 2022, \apj, 925, 79, \dodoi{10.3847/1538-4357/ac397d}

\bibitem[{{Lenz} \& {Breger}(2005)}]{2005CoAst.146...53L}
{Lenz}, P., \& {Breger}, M. 2005, Communications in Asteroseismology, 146, 53, \dodoi{10.1553/cia146s53}

\bibitem[{{L{\'e}pine} \& {Moffat}(1999)}]{1999ApJ...514..909L}
{L{\'e}pine}, S., \& {Moffat}, A. F.~J. 1999, \apj, 514, 909, \dodoi{10.1086/306958}

\bibitem[{{L{\'e}pine} {et~al.}(2000){L{\'e}pine}, {Moffat}, {St-Louis}, {Marchenko}, {Dalton}, {Crowther}, {Smith}, {Willis}, {Antokhin}, \& {Tovmassian}}]{2000AJ....120.3201L}
{L{\'e}pine}, S., {Moffat}, A. F.~J., {St-Louis}, N., {et~al.} 2000, \aj, 120, 3201, \dodoi{10.1086/316858}

\bibitem[{Levenberg(1944)}]{Levenberg1944AMF}
Levenberg, K. 1944, Quarterly of Applied Mathematics, 2, 164.
\newblock \url{https://api.semanticscholar.org/CorpusID:124308544}

\bibitem[{Li {et~al.}(2018)Li, Bedding, Murphy, Van~Reeth, Antoci, \& Ouazzani}]{10.1093/mnras/sty2743}
Li, G., Bedding, T.~R., Murphy, S.~J., {et~al.} 2018, Monthly Notices of the Royal Astronomical Society, 482, 1757, \dodoi{10.1093/mnras/sty2743}

\bibitem[{{Lightkurve Collaboration} {et~al.}(2018){Lightkurve Collaboration}, {Cardoso}, {Hedges}, {Gully-Santiago}, {Saunders}, {Cody}, {Barclay}, {Hall}, {Sagear}, {Turtelboom}, {Zhang}, {Tzanidakis}, {Mighell}, {Coughlin}, {Bell}, {Berta-Thompson}, {Williams}, {Dotson}, \& {Barentsen}}]{2018ascl.soft12013L}
{Lightkurve Collaboration}, {Cardoso}, J.~V.~d.~M., {Hedges}, C., {et~al.} 2018, {Lightkurve: Kepler and TESS time series analysis in Python}, Astrophysics Source Code Library.
\newblock \doeprint{1812.013}

\bibitem[{Marquardt(1963)}]{doi:10.1137/0111030}
Marquardt, D.~W. 1963, Journal of the Society for Industrial and Applied Mathematics, 11, 431

\bibitem[{{Massey} \& {Gronwall}(1990)}]{1990ApJ...358..344M}
{Massey}, P., \& {Gronwall}, C. 1990, \apj, 358, 344, \dodoi{10.1086/168991}

\bibitem[{{Moffat} {et~al.}(1988){Moffat}, {Drissen}, {Lamontagne}, \& {Robert}}]{1988ApJ...334.1038M}
{Moffat}, A. F.~J., {Drissen}, L., {Lamontagne}, R., \& {Robert}, C. 1988, \apj, 334, 1038, \dodoi{10.1086/166895}

\bibitem[{{Naz{\'e}} {et~al.}(2021){Naz{\'e}}, {Rauw}, \& {Gosset}}]{2021MNRAS.502.5038N}
{Naz{\'e}}, Y., {Rauw}, G., \& {Gosset}, E. 2021, \mnras, 502, 5038, \dodoi{10.1093/mnras/stab133}

\bibitem[{{Owocki} {et~al.}(1988){Owocki}, {Castor}, \& {Rybicki}}]{1988ApJ...335..914O}
{Owocki}, S.~P., {Castor}, J.~I., \& {Rybicki}, G.~B. 1988, \apj, 335, 914, \dodoi{10.1086/166977}

\bibitem[{Rate \& Crowther(2020)}]{10.1093/mnras/stz3614}
Rate, G., \& Crowther, P.~A. 2020, Monthly Notices of the Royal Astronomical Society, 493, 1512, \dodoi{10.1093/mnras/stz3614}

\bibitem[{Ricker {et~al.}(2014)Ricker, Winn, Vanderspek, Latham, Bakos, Bean, Berta-Thompson, Brown, Buchhave, Butler, Butler, Chaplin, Charbonneau, Christensen-Dalsgaard, Clampin, Deming, Doty, Lee, Dressing, Dunham, Endl, Fressin, Ge, Henning, Holman, Howard, Ida, Jenkins, Jernigan, Johnson, Kaltenegger, Kawai, Kjeldsen, Laughlin, Levine, Lin, Lissauer, MacQueen, Marcy, McCullough, Morton, Narita, Paegert, Palle, Pepe, Pepper, Quirrenbach, Rinehart, Sasselov, Sato, Seager, Sozzetti, Stassun, Sullivan, Szentgyorgyi, Torres, Udry, \& Villasenor}]{10.1117/1.JATIS.1.1.014003}
Ricker, G.~R., Winn, J.~N., Vanderspek, R., {et~al.} 2014, Journal of Astronomical Telescopes, Instruments, and Systems, 1, 014003, \dodoi{10.1117/1.JATIS.1.1.014003}

\bibitem[{{Saio}(2009)}]{2009CoAst.158..245S}
{Saio}, H. 2009, Communications in Asteroseismology, 158, 245

\bibitem[{{Sander} {et~al.}(2012){Sander}, {Hamann}, \& {Todt}}]{2012A&A...540A.144S}
{Sander}, A., {Hamann}, W.~R., \& {Todt}, H. 2012, \aap, 540, A144, \dodoi{10.1051/0004-6361/201117830}

\bibitem[{{Sander} {et~al.}(2019){Sander}, {Hamann}, {Todt}, {Hainich}, {Shenar}, {Ramachandran}, \& {Oskinova}}]{2019A&A...621A..92S}
{Sander}, A.~A.~C., {Hamann}, W.~R., {Todt}, H., {et~al.} 2019, \aap, 621, A92, \dodoi{10.1051/0004-6361/201833712}

\bibitem[{{Sander} {et~al.}(2020){Sander}, {Vink}, \& {Hamann}}]{2020MNRAS.491.4406S}
{Sander}, A. A.~C., {Vink}, J.~S., \& {Hamann}, W.~R. 2020, \mnras, 491, 4406, \dodoi{10.1093/mnras/stz3064}

\bibitem[{{Skrutskie} {et~al.}(2006){Skrutskie}, {Cutri}, {Stiening}, {Weinberg}, {Schneider}, {Carpenter}, {Beichman}, {Capps}, {Chester}, {Elias}, {Huchra}, {Liebert}, {Lonsdale}, {Monet}, {Price}, {Seitzer}, {Jarrett}, {Kirkpatrick}, {Gizis}, {Howard}, {Evans}, {Fowler}, {Fullmer}, {Hurt}, {Light}, {Kopan}, {Marsh}, {McCallon}, {Tam}, {Van Dyk}, \& {Wheelock}}]{2006AJ....131.1163S}
{Skrutskie}, M.~F., {Cutri}, R.~M., {Stiening}, R., {et~al.} 2006, \aj, 131, 1163

\bibitem[{Smith {et~al.}(2012)Smith, Stumpe, Cleve, Jenkins, Barclay, Fanelli, Girouard, Kolodziejczak, McCauliff, Morris, \& Twicken}]{Smith_2012}
Smith, J.~C., Stumpe, M.~C., Cleve, J. E.~V., {et~al.} 2012, Publications of the Astronomical Society of the Pacific, 124, 1000, \dodoi{10.1086/667697}

\bibitem[{{St-Louis} {et~al.}(2009){St-Louis}, {Chen{\'e}}, {Schnurr}, \& {Nicol}}]{2009ApJ...698.1951S}
{St-Louis}, N., {Chen{\'e}}, A.~N., {Schnurr}, O., \& {Nicol}, M.~H. 2009, \apj, 698, 1951, \dodoi{10.1088/0004-637X/698/2/1951}

\bibitem[{{St-Louis} {et~al.}(2020){St-Louis}, {Piaulet}, {Richardson}, {Shenar}, {Moffat}, {Eversberg}, {Hill}, {Gauza}, {Knapen}, {Kub{\'a}t}, {Kub{\'a}tov{\'a}}, {Sablowski}, {Sim{\'o}n-D{\'\i}az}, {Bolduan}, {Dias}, {Dubreuil}, {Fuchs}, {Garrel}, {Grutzeck}, {Hunger}, {K{\"u}sters}, {Langenbrink}, {Leadbeater}, {Li}, {Lopez}, {Mauclaire}, {Moldenhawer}, {Potter}, {dos Santos}, {Schanne}, {Schmidt}, {Sieske}, {Strachan}, {Stinner}, {Stinner}, {Stober}, {Strandbaek}, {Syder}, {Verilhac}, {Waldschl{\"a}ger}, {Weiss}, \& {Wendt}}]{2020MNRAS.497.4448S}
{St-Louis}, N., {Piaulet}, C., {Richardson}, N.~D., {et~al.} 2020, \mnras, 497, 4448, \dodoi{10.1093/mnras/staa2214}

\bibitem[{Stumpe {et~al.}(2012)Stumpe, Smith, Cleve, Twicken, Barclay, Fanelli, Girouard, Jenkins, Kolodziejczak, McCauliff, \& Morris}]{Stumpe_2012}
Stumpe, M.~C., Smith, J.~C., Cleve, J. E.~V., {et~al.} 2012, Publications of the Astronomical Society of the Pacific, 124, 985, \dodoi{10.1086/667698}

\bibitem[{{Sundqvist} {et~al.}(2018){Sundqvist}, {Owocki}, \& {Puls}}]{2018A&A...611A..17S}
{Sundqvist}, J.~O., {Owocki}, S.~P., \& {Puls}, J. 2018, \aap, 611, A17, \dodoi{10.1051/0004-6361/201731718}

\bibitem[{{Toal{\'a}} {et~al.}(2022){Toal{\'a}}, {Bowman}, {Van Reeth}, {Todt}, {Dsilva}, {Shenar}, {Koenigsberger}, {Estrada-Dorado}, {Oskinova}, \& {Hamann}}]{2022MNRAS.514.2269T}
{Toal{\'a}}, J.~A., {Bowman}, D.~M., {Van Reeth}, T., {et~al.} 2022, \mnras, 514, 2269, \dodoi{10.1093/mnras/stac1455}

\bibitem[{{Tody}(1993)}]{1993ASPC...52..173T}
{Tody}, D. 1993, in Astronomical Society of the Pacific Conference Series, Vol.~52, Astronomical Data Analysis Software and Systems II, ed. R.~J. {Hanisch}, R.~J.~V. {Brissenden}, \& J.~{Barnes}, 173

\bibitem[{{Townsend} \& {MacDonald}(2006)}]{2006MNRAS.368L..57T}
{Townsend}, R.~H.~D., \& {MacDonald}, J. 2006, \mnras, 368, L57, \dodoi{10.1111/j.1745-3933.2006.00157.x}

\bibitem[{{Vollmann} \& {Eversberg}(2006)}]{2006AN....327..862V}
{Vollmann}, K., \& {Eversberg}, T. 2006, Astronomische Nachrichten, 327, 862, \dodoi{10.1002/asna.200610645}

\bibitem[{{Zechmeister} \& {K{\"u}rster}(2009)}]{2009A&A...496..577Z}
{Zechmeister}, M., \& {K{\"u}rster}, M. 2009, \aap, 496, 577, \dodoi{10.1051/0004-6361:200811296}

\end{thebibliography}
\bibliographystyle{aasjournal}

\end{document}